\documentclass{article}

\usepackage[english]{babel}

\usepackage[letterpaper,top=2cm,bottom=2cm,left=3cm,right=3cm,marginparwidth=1.75cm]{geometry}

\usepackage{amsmath}
\usepackage{graphicx}
\usepackage[colorlinks=true, allcolors=blue]{hyperref}
\usepackage{amsmath}
\usepackage{graphicx}
\usepackage[colorlinks=true, allcolors=blue]{hyperref}
\usepackage{amsthm}
\usepackage{tikz}
\usepackage{centernot}
\usepackage{mathtools}
\usepackage{ stmaryrd }
\usetikzlibrary{matrix,arrows.meta}
\usepackage{amsmath}
\usepackage{tikz}
\usetikzlibrary{matrix,arrows,decorations.pathmorphing}
\usetikzlibrary{shapes,arrows,positioning}
\usepackage{graphicx}
\usepackage{amsmath}
\usepackage{amssymb}
\usepackage{latexsym}
\usepackage{amssymb,amsmath}
\usepackage{tikz-cd}
\usetikzlibrary{arrows}
\usepackage{amsfonts}
\usepackage{amssymb}
\usepackage{graphicx}
\usepackage{fancyhdr}
\usepackage{fancyvrb}
\usepackage{fancybox}
\usepackage{float}
\usepackage{geometry}
\usepackage{minitoc}
\usepackage{indentfirst}
\usepackage{multicol, color}
\usepackage[outerbars]{changebar}
\usepackage{pifont,textcomp}
\usepackage{amsfonts}
\usepackage{amsmath}
\usepackage{amscd}
\usepackage{array}
\usepackage{multirow}
\usepackage{mathrsfs}
\usepackage{amssymb}
\usepackage{amsthm}
\usepackage{amsmath}
\usepackage{subfig}
\usepackage{graphicx}
\usepackage{booktabs}
\usepackage[colorlinks=true, allcolors=blue]{hyperref}

\title{Rényi and Tsallis information entropies for a  harmonic position-dependent mass}

\author{ Dorcas A. Addo$^{1}$\footnote{Email:daaddo@uew.edu.gh},\,\, Daniel Sabi-Takou$^{2,3}$\footnote{Email: sabitakoudaniel11@gmail.com},\,\,  Eugene Adjei$^{4}$\footnote{Email:eyaadjei@ug.edu.gh},\,\, Latévi M. Lawson$^{5}$\footnote{ Email:latevi@aims.edu.gh\, (Corresponding author)}
 \\\\
 $^{1}$ Department of Mathematics Education, University of Education, Winneba\\ P. O. Box 25, Winneba, Ghana\\\\
$^{2}$ Ecole Polytechnique d'Abomey Calavi (EPAC-UAC),\\
Universit\'e d'Abomey-Calavi (UAC), B\'enin\\
 $^{3}$  Unit\'e de Recherche en Physique Th\'eorique (URPT),\\
 Institut de Math\'ematiques et de Sciences  Physiques (IMSP),\\
 01 B.P. 613 Porto-Novo, Rep. du B\'enin\\\\
 $^{4}$Department of Mathematics,
 University of Ghana,\\ Legon, GA-489-9348,\,P.O.Box LG 62, Accra, Ghana\\\\
 $^{5}$African Institute for Mathematical Sciences (AIMS),\\ 1 Shoppers Street, Manet, Spintex, Accra, Ghana.\\
   }

\begin{document}
\maketitle

\begin{abstract}
	In this paper, we study Rényi and Tsallis information entropies for a Hamiltonian system with 	position-dependent mass confined in harmonic oscillator potential. Gegenbauer polynomials are 	used to obtain the position eigenfunction of such a system, and the modified Bessel function of the	second kind is used to determine the equivalent momentum eigenfunction. By means of probability densities of both representations, we analytically and numerically evaluate the Heisenberg-like 	uncertainty of this system. Because the Rényi and Tsallis information entropies in position representation are described by integral functionals of the Gegenbauer polynomials, these quantities 	are much more difficult to calculate. To get around this problem, we evaluate these information entropies at the system’s asymptotical limit, which corresponds to the behavior of an undeformed 	harmonic oscillator. Nevertheless, no approximation method is used to obtain the Tsallis and 	Rényi information entropies in momentum representation. In both representations, we find that 	these information entropies approach the Shannon entropy when the entropic parameter $\alpha\rightarrow 1$	and are closed to the results of similar models of the literature. Finally, we evaluate the related 	entropic uncertainty relations numerically to validate the the latter observations.

\end{abstract}

\section{Introduction}
The study of information entropy \cite{1,2,3,4,5,6}  has become an important tool for understanding the localization, uncertainty, and statistical structure of quantum systems \cite{7,8,9,10,11} . Among the many entropy measures proposed in quantum information theory, the Rényi \cite{2} and Tsallis\cite{3} entropies are especially useful because they generalize the Shannon entropy \cite{1} and provide a tunable parameter that highlights different aspects of a probability distribution. Rényi and Tsallis entropies have found applications in different disciplines of sciences such as: in coding theory \cite{13}, in  high-energy collisions
\cite{14,15}, in fractional diffusion processes \cite{16}, in conformal field theory \cite{17,18}, in  black hole area law \cite{19}  and in many many others (see, for example, sources in \cite{20,21,22,23,24,25}).

In the last few decades, position-dependent mass (PDM) systems \cite{27} have attracted considerable attention because they  naturally arise  in the description  of semiconductor  and their relation to optical and electronic properties \cite{28,29,30}. In gravitational physics, PDM models have been studied in cosmological settings \cite{31,32} and in the context of quantum gravity \cite{33,34,35,36,37,37a,37b}. The majority of such studies dedicated to the problems relevance to condensed matter and solid state physics \cite{38,39}. A particular interesting example  is a  harmonic oscillator with position-dependent mass \cite{40,41,42}, since it combines one of the most fundamental exactly solvable models in quantum mechanics with the additional complexity introduced by mass variation.  This makes it a valuable testbed for exploring how spatial inhomogeneity modifies the quantum properties of bound states. 

As we have seen, Rényi and Tsallis information entropies  and PDM systems have been  extensively studied in the literature; however, as far as we know,  the study   of information entropies  for PDM systems    has received less  attention  \cite{42b,
43,44, 45}. In this work, we analyse how Rényi and Tsallis entropies provide information about the localization properties and uncertainty structure of  PDM confined in confined in harmonic oscillator potential.
    
We first calculate the position and momentum space variances and the associated Heisenberg-like uncertainty using the eigenfunctions of this system, which are expressed in terms of Gegenbauer polynomials in position representation and the modified Bessel function of the second kind in momentum representation. The determination of the  Rényi and Tsallis information entropies in position representation are  far more difficult \cite{46,47,48}. This is because these quantities are
described by means of some power or logarithmic functionals of the position probability density (expressed in terms of Gegenbauer polynomial), which cannot be calculated in an analytical way nor numerically computed \cite{46,47,48}.  Therefore, in the second part of this study, we analytically and numerically evaluate the asymptotical behaviors of the Rényi and Tsallis entropies using  the modern techniques of approximation theory \cite{48,49,49a,49b,49c,49d}.
The results are comparable to those obtained lately for the undeformed harmonic oscillator \cite{44, 49e}.
However, the Rényi and Tsallis entropies in momentum space controlled by the  modified Bessel function of the second kind  are obtained without any approximation theory. Since the system at hand manifests an ordinary harmonic oscillator behavior at the asymptotic limit, we recover the  Rényi and Tsallis entropy behavior of harmonic oscillor \cite{49e} in momentum representation  closed to the asymoptical limit of this system. Finally, with the obtained results in both represenations,  we evaluate  the  Rényi and Tsallis entropy-based uncertainty principle  for this system.

The paper is organized as follows: In the next section, we review   the dynamics of PDM in a harmonic oscillator potential \cite{40}. By means of the probabilty densities in position and momentum representations of this system, we show that the Heisenberg like uncertainty relation is satisfied. Section \eqref{sec3} presents  in one hand,  analytical and numerical expressions for the entropic moments, the  Rényi and Tsallis entropies  in position and momentum representation. In the second hand, we show that entropic
uncertainty relations are always satisfied. Finally, a concluding Section\eqref{sec4} closes the paper.

 \section{Harmonic position-dependent mass}
In this section, we briefly review the spectrum of  position-dependent mass system confined in harmonic oscillator potential had recently been obtained in \cite{40}. From the system eigenfunction,  we
give the  position and momentum probability densities which we need for the rest of this work. Finally, we  calculate position and momentum Heisenberg-like based uncertainty relations.

\subsection{Spectrum of the system}
\label{sec2}

The most general Hermitian kinetic Hamiltonian for a particle with position–dependent mass $m(\hat x)$ can be written as \cite{27}

\begin{equation}
\hat T=
\frac{1}{4}
\left[
m^\nu(\hat x)\hat p\, m^\mu(\hat x)\hat p\, m^\gamma(\hat x)
+
m^\gamma(\hat x)\hat p\, m^\mu(\hat x)\hat p\, m^\nu(\hat x)
\right],
\end{equation}
where $\nu$, $\mu$ and $\gamma$ are ordering parameters constrained by
$\nu+\mu+\gamma=-1$ . Clearly, there are different Hamiltonians depending on the choices of the parameters \cite{51,52,53,54}. Here we shall work with the Mustafa and Mazharimousavi form \cite{54} which corresponds to the choice  
$\nu=\gamma=-\frac{1}{4}$ and  $\mu=-\frac{1}{2}$, 
\begin{equation}
\hat T=
\frac{1}{2}
\frac{1}{m^{1/4}(\hat x)}
\hat p
\frac{1}{m^{1/2}(\hat x)}
\hat p
\frac{1}{m^{1/4}(\hat x)} .
\end{equation}
The Hamiltonian of the system is therefore written as

\begin{equation}
\hat H=
\hat T+V(\hat x)
=
\frac{1}{2}
\frac{1}{m^{1/4}(\hat x)}
\hat p
\frac{1}{m^{1/2}(\hat x)}
\hat p
\frac{1}{m^{1/4}(\hat x)}
+
V(\hat x),
\end{equation}
where $V(\hat x)$ denotes the potential energy. The time-independent Schr\"{o}dinger equation  for  a harmonic oscillator particle, i.e. $ V( x)=\frac{1}{2}m_0\omega^2 x^2$  is given by 
\begin{eqnarray}\label{a}
E\phi(x)=-\frac{\hbar^2}{2m_0}\sqrt[4]{\frac{m_0}{m(x)}}\frac{d}{dx}\sqrt{\frac{m_0}{m( x)}}\frac{d}{dx}\sqrt[4]{\frac{m_0}{m( x)}}\phi(x)+V(x)\phi(x),
\end{eqnarray}
where $m_0$ is a constant mass,  $E$ is the energy spectrum, $\phi(x)$ is the wavefunction defined on a Hilbert space $\mathcal{H}=\mathcal{L}^2(\mathbb{R})$.  For the present analysis, we adopted in \cite{40} the following mass distribution
\begin{equation}
m(x)=\frac{m_0}{(1+\tau x^2)^2},
\label{71}
\end{equation}
where $\tau$ is a deformation parameter satisfying $0<\tau<1$.  Physically, the parameter $\tau$ may be interpreted as an inverse square length scale characterizing structural inhomogeneities such as curvature effects or defects in the medium.

Solving Eq.~(\ref{a}) leads to the following energy spectrum

\begin{equation} 
E_n=
\hbar\omega\left(n+\frac{1}{2}\right)
\sqrt{1+\frac{\tau^2\hbar^2}{4m_0^2\omega^2}}
+
\frac{\tau\hbar^2}{2m_0}
\left(n^2+2n+\frac{1}{2}\right),
\label{en11}
\end{equation}
while the normalized eigenfunctions are given by
\begin{equation}
\phi_n^\lambda(x)=
N_n
\left(\frac{1}{1+\tau x^2}\right)^{\frac{\lambda+1}{2}}
C_n^{(\lambda)}
\left(
\frac{x\sqrt{\tau}}{\sqrt{1+\tau x^2}}
\right),
\label{di}
\end{equation}
where  $C_n^{(\lambda)}$ is 
the constant $\lambda$ and the normalization constant read
\begin{equation}
\lambda=
\frac{1}{2}+
\frac{1}{2}
\sqrt{1+4\frac{m_0^2\omega^2}{\tau^2\hbar^2}},\quad N_n=
\sqrt{
\frac{n!(n+\lambda)[\Gamma(\lambda)]^2}
{\pi 2^{1-2\lambda}\Gamma(n+2\lambda)}
}.
\end{equation}
At the limit $n=0$, we find the spectrum of the ground states of the system
\begin{eqnarray}
    \lim_{n\rightarrow 0}E_n&=& E_0= \frac{1}{2}\hbar\omega
\sqrt{1+\frac{\tau^2\hbar^2}{4m_0^2\omega^2}}
+\frac{\tau\hbar^2}{4m_0},\\
    \lim_{n\rightarrow 0}\phi_n(x)&=&\phi_0(x)= N_0
\left(\frac{1}{1+\tau x^2}\right)^{\frac{\lambda+1}{2}},
\end{eqnarray}
where $C_n^{(\lambda)}(y)=1$. 
 However, at limit $\tau\rightarrow 0$,  we have $\lambda \rightarrow \infty$, we find that the wavefunctions \eqref{di}  and  the eigenvalues \eqref{en11} converge to the spectrum of an ordinary harmonic oscillator 
\begin{eqnarray}
\lim_{\tau\rightarrow 0}E_n&=&\varepsilon_n=\hbar\omega\left(n+\frac{1}{2}\right),\\
 \lim_{\tau\rightarrow 0} \phi_n^\lambda(x)&=&\phi_n^{\infty}(x)
=\frac{1}{\sqrt{2^nn!}} \left(\frac{m_0\omega}{\pi\hbar}\right)^{1/4} e^{-\frac{m_0\omega }{2\hbar}x^2}
H_n\left(\sqrt{\frac{m_0\omega}{\hbar}}\,x\right),\label{qa}
    \end{eqnarray}
where $H_n(x)$  is the Hermite polynomial. The reader can check equation \eqref{qa} from appendix \eqref{xv}.

\begin{figure}
\centering
\includegraphics[width=6cm, height=5cm]{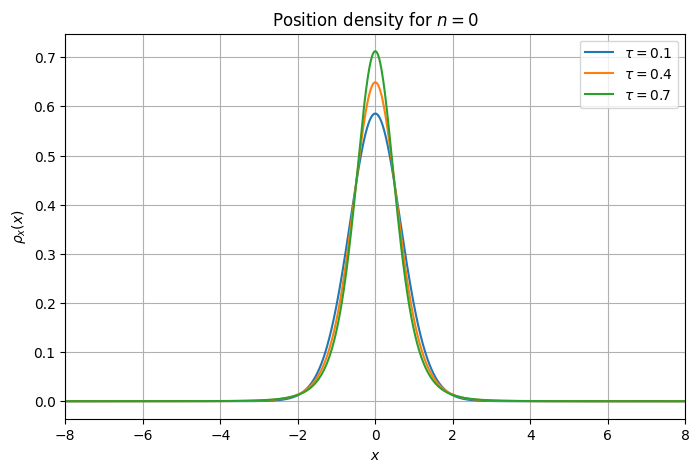}
\includegraphics[width=6cm, height=5cm]{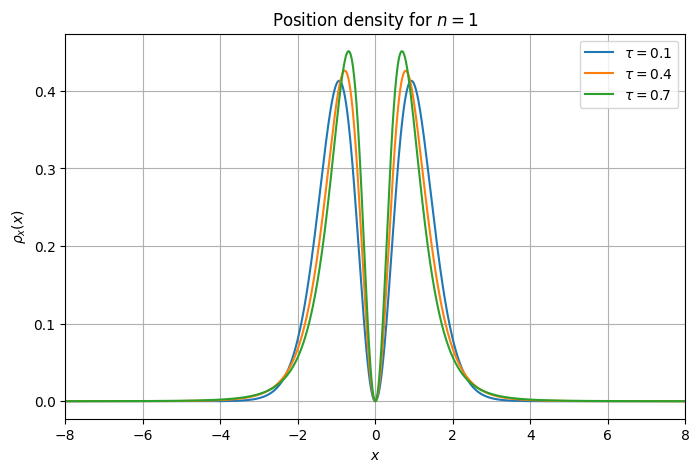}
\includegraphics[width=6cm, height=5cm]{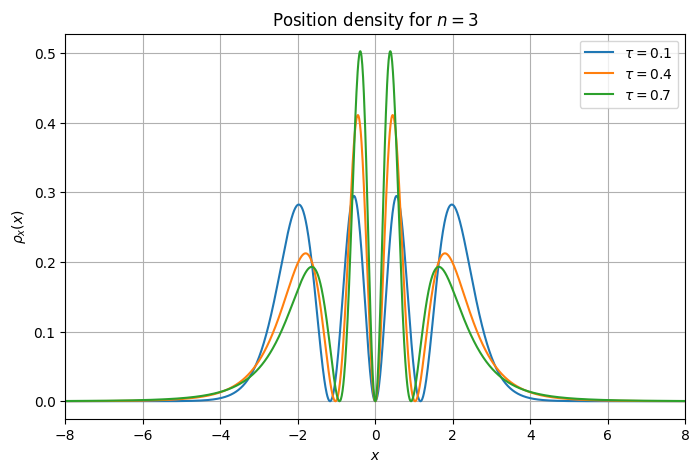}
\includegraphics[width=6cm, height=5cm]{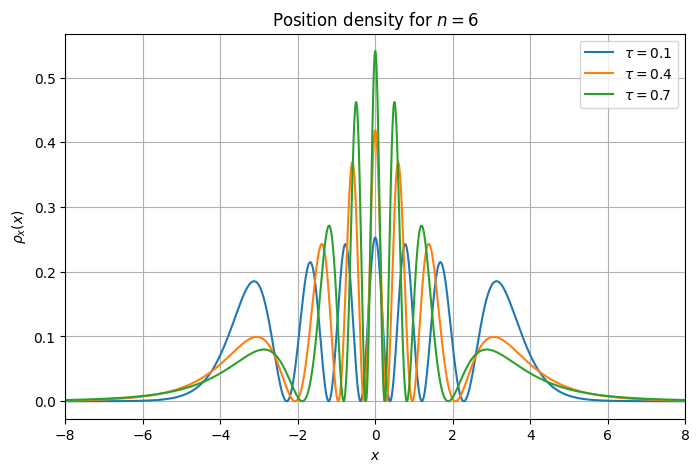}
\caption{Probability density function in position space for fixed values of parameter $\tau$ and quantum number $n$.}
\label{fg1}
\end{figure}

The probability density in position space is given by
\begin{equation}\label{p1}
\rho_n^\lambda(x)=|\phi_n^\lambda(x)|^2
=
N_n^2
\left(\frac{1}{1+\tau x^2}\right)^{\lambda+1}
\left|
C_n^{(\lambda)}
\left(
\frac{x\sqrt{\tau}}{\sqrt{1+\tau x^2}}
\right)
\right|^2 .
\end{equation}
Figure\,\ref{fg1} depicts probability density plots for the states
$n = 0; n = 1; n = 3; n = 6$ for some fixed values of $\tau=0.1;\, 0.4;\, 0.7$. It can be seen that the probability of finding a
particle  is increasing with the quantum numbers  and the for the parameters of deformation when $x\in [-2,2]$. However, this probability is discreasing when the  position tends to the infinities. 

On the other hand, the Fourier transform of the position eigenfunction $ \phi_n^\lambda (x)$ given by (\ref{di})
provides the eigenfunction of the system in the conjugated momentum space as
\begin{eqnarray} \label{fourier}
\phi_n^\lambda(p)=
\frac{1}{\sqrt{2\pi\hbar}}
\int_{-\infty}^{+\infty}
e^{-ipx}\phi_n^\lambda(x)\,dx.
\end{eqnarray}

A change of variable from $x$ to
\begin{equation}\label{wt}
y=\frac{x\sqrt{\tau}}{\sqrt{1+\tau x^2}},
\end{equation}
leads to the integral representation
\begin{equation}\label{qw}
\phi_n^\lambda(p)=
\frac{N_n}{\sqrt{2\pi\hbar\tau}}
\int_{-1}^{+1}
e^{-ip\frac{y}{\sqrt{\tau(1-y^2)}}}
(1-y^2)^{\frac{\lambda-2}{2}}
C_n^{(\lambda)}(y)\,dy .
\end{equation}
Using the hypergeometric and the  Pochhammer symbol representations of the Gegenbauer polynomials \cite{55}
\begin{eqnarray}
C_n^\lambda(y)&=& \frac{(2\lambda)_n}{n!} {}_2F_1 \left(-\frac{n}{2}, \frac{n}{2}+\lambda; \lambda+\frac{1}{2}; 1-y^2\right)\cr 
  &=& \frac{(2\lambda)_n}{n!} \sum_{k=0}^{n} \frac{(-\frac{n}{2})_k (\frac{n}{2}+\lambda)_k}{(\lambda+\frac{1}{2})_k}\frac{(1-y^2)^k}{k!},
\end{eqnarray}
the  equation \eqref{qw} becomes
\begin{eqnarray}
\phi_n^\lambda(p)=
\frac{N_n}{\sqrt{2\hbar\pi\tau}}  \frac{(2\lambda)_n}{n!} \sum_{k=0}^{n} \frac{(-\frac{n}{2})_k (\frac{n}{2}+\lambda)_k }{(\lambda+\frac{1}{2})_k k!}
\int_{-1}^{+1}
e^{-ip\frac{y}{\sqrt{\tau(1-y^2)}}}
(1-y^2)^{\frac{\lambda}{2}+k-1}
\,dy,
\end{eqnarray}
where  $(z)_a = \frac{\Gamma(z +a)}{\Gamma(z)}$ is the Pochhammer symbol. The last integral can be evaluated in terms of the modified Bessel function of the second kind $K_\nu(x)$ \cite{55}, giving
\begin{eqnarray}
   \int_{-1}^{+1}
e^{-ip\frac{y}{\sqrt{\tau(1-y^2)}}}
(1-y^2)^{\frac{\lambda}{2}+k-1}
dy 
= \frac{2\sqrt{\pi}}{\Gamma\!\left(\frac{\lambda+1}{2}+k\right)}
\left(\frac{|p|}{2\sqrt{\tau}}\right)^{\frac{\lambda}{2}+k}
K_{\frac{\lambda}{2}+k}\!\left(\frac{|p|}{\sqrt{\tau}}\right).
\end{eqnarray}
 Finally we have
\begin{eqnarray}
   \phi_n^\lambda(p)&=& \sqrt{2\pi}\frac{N_n}{\sqrt{\hbar \tau}} \frac{\Gamma(2\lambda+n)}{\Gamma(2\lambda)n!}\frac{\Gamma\left(\lambda+\frac{1}{2}\right)}{\Gamma(-\frac{n}{2})\Gamma(\frac{n}{2}+\lambda)}\sum_{k=0}^n\frac{\Gamma(-\frac{n}{2}+k)\Gamma(\frac{n}{2}+\lambda+k)}{\Gamma\left(\frac{\lambda+1}{2}+k\right)\Gamma\left(\lambda+\frac{1}{2}+k\right)k!}\cr&&\times \left(\frac{|p|}{2\sqrt{\tau}}\right)^{\frac{\lambda}{2}+k}K_{\frac{\lambda}{2}+k}\!\left(\frac{|p|}{\sqrt{\tau}}\right).
\end{eqnarray}
Consequently, the probability density in momentum space $ \rho_n^\lambda(p)=|\phi_n^\lambda(p)|^2$ is defined as
\begin{eqnarray}
\rho_n^\lambda(p)= A_n^2 \left(\frac{|p|}{2\sqrt{\tau}}\right)^{\lambda+2k}K_{\frac{\lambda}{2}+k}^2\!\left(\frac{|p|}{\sqrt{\tau}}\right),
\end{eqnarray}
where 
\begin{eqnarray}
A_{n,k}^2=  \frac{2\pi N_n^2}{\hbar \tau}\left(\frac{\Gamma(2\lambda+n)}{\Gamma(2\lambda)n!}\frac{\Gamma\left(\lambda+\frac{1}{2}\right)}{\Gamma(-\frac{n}{2})\Gamma(\frac{n}{2}+\lambda)}\right)^2\sum_{k=0}^n\left(
\frac{\Gamma(-\frac{n}{2}+k)\Gamma(\frac{n}{2}+\lambda+k)}{\Gamma(-\frac{n}{2}+k)\Gamma\left(\frac{\lambda+1}{2}+k\right)\Gamma\left(\lambda+\frac{1}{2}+k\right)k!}
\right)^2.
\end{eqnarray}

\begin{figure}
\centering
\includegraphics[width=6cm, height=5cm]{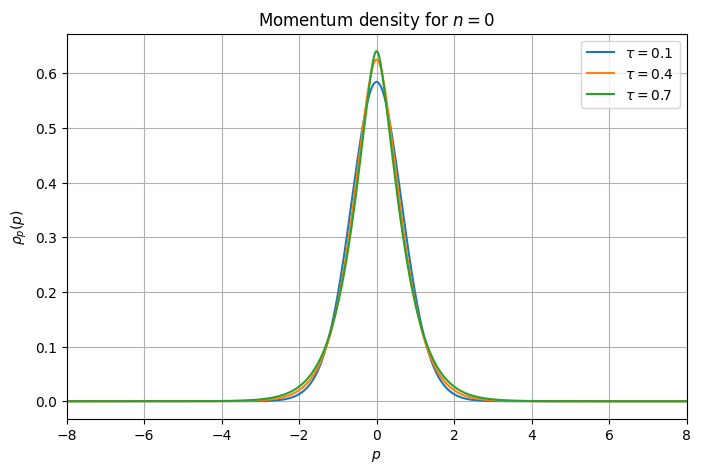}
\includegraphics[width=6cm, height=5cm]{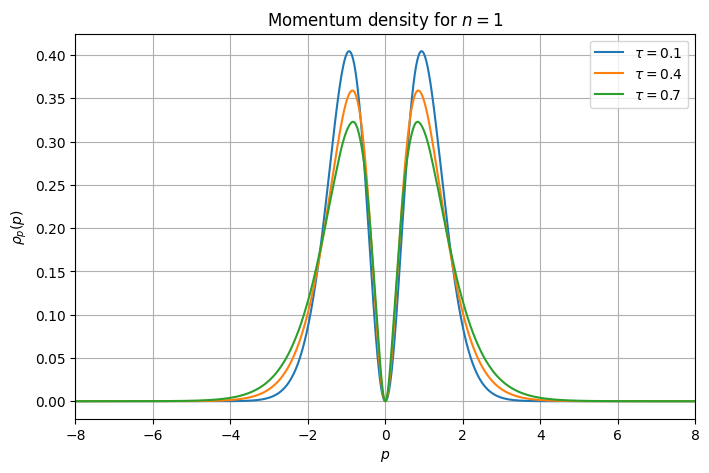}
\includegraphics[width=6cm, height=5cm]{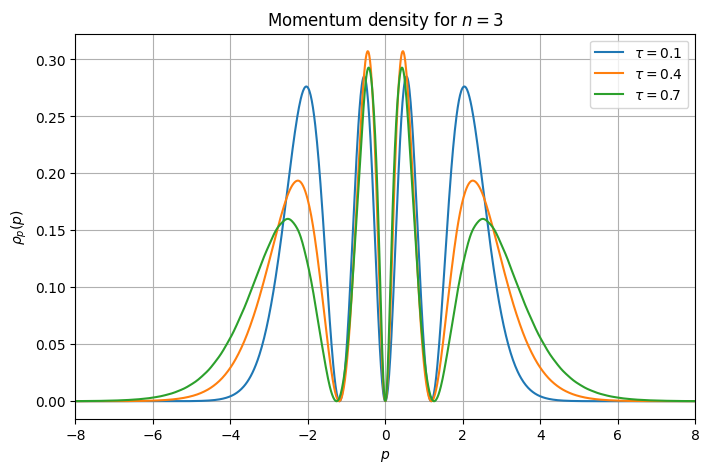}
\includegraphics[width=6cm, height=5cm]{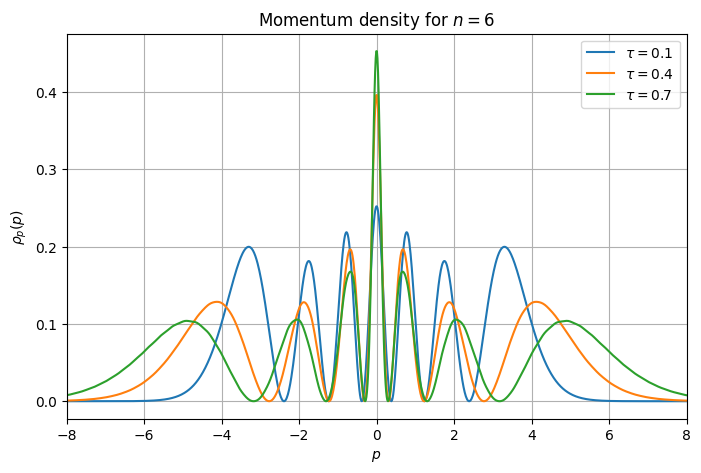}
\caption{Density function in momentum space for fixed values of parameter $\tau$ and quantum number $n$.}
\label{fig2}
\end{figure}
 Figure \eqref{fig2} depicts momentum probability density plots for the states
$n = 0; n = 1; n = 3; n = 6$ for some fixed values of $\tau=0.1;\, 0.4;\, 0.7$. It can be seen that the probability of finding a
particle  is increasing with the quantum numbers when $p\in [-2,2]$,  but is  discreasing  when  the parameters are increasing when the values of the momentum goes to the infinities.

\subsection{Heisenberg-like uncertainty}
For the system at hand, the position-space variance $\sigma_x^2=\langle \hat x^2\rangle-\langle \hat x\rangle^2$ and  the corresponding momentum-space 
variance is given by $\sigma_p^2=\langle p^2\rangle-\langle p\rangle^2$. The generalized Heisenberg (or Heisenberg-like) uncertainty   of this  system is given by
\begin{eqnarray}\label{Hei}
    \sigma_x^2\sigma_p^2\geq \frac{\hbar^2}{4}.
\end{eqnarray}

The position expectation values of the  system with the  density $ \rho_n^\lambda(x)$, are defined by

\begin{eqnarray}
  \langle \hat x\rangle&=&  \int_{-\infty}^{+\infty} x \rho_n^\lambda(x)dx= N_n^2\int_{-\infty}^{+\infty} x
\left(\frac{1}{1+\tau x^2}\right)^{\lambda+1}
\left|
C_n^{(\lambda)}
\left(
\frac{x\sqrt{\tau}}{\sqrt{1+\tau x^2}}
\right)
\right|^2dx,\\
 \langle \hat x^2\rangle&=&  \int_{-\infty}^{+\infty} x^2 \rho_n^\lambda(x)dx= N_n^2\int_{-\infty}^{+\infty} x^2
\left(\frac{1}{1+\tau x^2}\right)^{\lambda+1}
\left|
C_n^{(\lambda)}
\left(
\frac{x\sqrt{\tau}}{\sqrt{1+\tau x^2}}
\right)
\right|^2dx.
\end{eqnarray}

Using the property of Gegenbauer polynomials \cite{55}
\begin{equation}
C_n^{(\lambda)}(-y)=(-1)^n C_n^{(\lambda)}(y)\implies
\left|C_n^{(\lambda)}(-y)\right|^2 = \left|C_n^{(\lambda)}(y)\right|^2,
\end{equation}
one can see that $\rho_n^\lambda(x)$ is an even function, and the integrand $x\,\rho_n^\lambda(x)$ is odd. Therefore, by symmetry,
\begin{equation}
\langle \hat x\rangle = 0.
\end{equation}
However, using equation \eqref{wt} we have
\begin{eqnarray} 
 \langle \hat x^2\rangle=\frac{ N_n^2}{\tau^{\frac{3}{2}}} \int_{-1}^{+1}y^2
\left(1-y^2\right)^{\lambda-\frac{3}{2}}
\left|C_n^{(\lambda)}
\left(y\right)
\right|^2dy.\label{2q}
\end{eqnarray}
The expectation $\langle \hat x^2\rangle$ of equations \eqref{2q} and the position variance $\sigma_x^2$  give 
\begin{eqnarray}
\sigma_x^2 = \langle \hat{x}^2 \rangle  = \frac{n(n+2\lambda)}{\tau\,(2\lambda-1)(2n+2\lambda-1)}
\end{eqnarray}

\begin{figure}
\centering
\includegraphics[width=7cm, height=6cm]{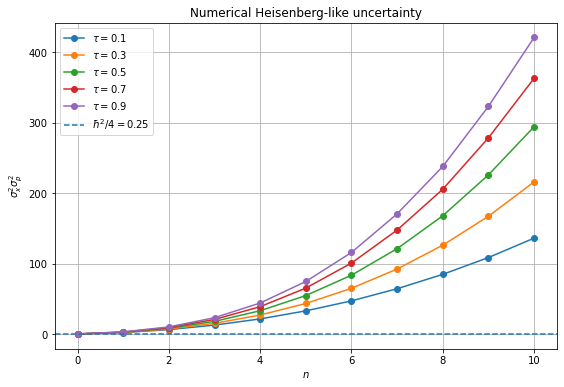}
\caption{ Heisenberg uncertainty \eqref{Hei} as  function of quantum number $n$  for fixed values of parameter $\tau$}
\label{fig3}
\end{figure}

Similarly, the corresponding momentum expectation values of the  system with the  density $ \rho_n^\lambda(x)$, are defined by

\begin{eqnarray}
  \langle \hat p\rangle=  \int_{-\infty}^{+\infty} p \rho_n^\lambda(p)dp,\quad \mbox{and} \quad 
 \langle \hat p^2\rangle=  \int_{-\infty}^{+\infty} p^2 \rho_n^\lambda(p)dp.
\end{eqnarray}
Since the probability density $\rho_n^\lambda(p)$ depends only on $|p|$, it is an even function:
\begin{equation}
\rho_n^\lambda(-p) = \rho_n^\lambda(p).
\end{equation}
However, the integrand $p\,\rho_n^\lambda(p)$ is an odd function since the function $p$ is odd. Consequently, by symmetry,
\begin{equation}
\langle \hat p\rangle = 0.
\end{equation}
The second moment expectation gives
\begin{equation}
\langle \hat p^2\rangle = \int_{-\infty}^{+\infty} p^2\, \rho_n^\lambda(p)\,dp=2 \int_{0}^{+\infty} p^2\, \rho_n^\lambda(p)\,dp.
\end{equation}
Substituting the explicit form of the density, we obtain
\begin{equation}
\langle \hat p^2\rangle
= A_{n,k}^2 2^{1-2k-\lambda} \tau^{k-\frac{\lambda}{2}} 
\int_{0}^{+\infty} 
p^{\lambda+2k+2}
K_{\frac{\lambda}{2}+k}^2\!\left(\frac{p}{\sqrt{\tau}}\right)dp.
\end{equation}
Performing the following change of variable
\begin{equation}\label{tr}
u = \frac{p}{\sqrt{\tau}}\quad \mbox{and}\quad dp=\sqrt{\tau}du,
\end{equation}
we obtain
\begin{equation}
\langle \hat p^2\rangle
=
2^{1-2k-\lambda} \tau^{2k+\frac{3}{2}} A_n^2
\int_{0}^{+\infty} 
u^{\lambda+2k+2}
K_{\frac{\lambda}{2}+k}^2(u)\,du.
\end{equation}
Using the standard integral formula for the square of the modified Bessel function \cite{55},
\begin{equation}
\int_0^\infty u^{\mu-1} K_\nu^2(u)\,du
=\frac{\sqrt{\pi}\,
\Gamma\!\left(\frac{\mu}{2}\right)
\Gamma\!\left(\frac{\mu}{2}+\nu\right)
\Gamma\!\left(\frac{\mu}{2}-\nu\right)}
{4\,\Gamma\!\left(\frac{\mu+1}{2}\right)},
\end{equation}
with $\mu = \lambda + 2k + 3$ and $\nu = \frac{\lambda}{2}+k$, we finally obtain
\begin{equation}
\langle \hat p^2\rangle=A_{n,k}^2
2^{-2k-\lambda-1}
\sqrt{\pi}\,
\tau^{2k+\frac{3}{2}}
\frac{
\Gamma\!\left(\frac{\lambda}{2}+k+\frac{3}{2}\right)
\Gamma\!\left(\lambda+2k+\frac{3}{2}\right)
\Gamma\!\left(\frac{3}{2}\right)
}
{\Gamma\!\left(\frac{\lambda}{2}+k+2\right)}.
\end{equation}
The symmetry of the momentum probability distribution is reflected by the vanishing of $\langle \hat p\rangle$, however the momentum fluctuation are characterized by the non-zero value $\langle \hat p^2\rangle$. In particular, the variance is given by
\begin{equation}
\sigma_p^2= \langle \hat p^2\rangle= A_{n,k}^2 2^{-2k-\lambda-1}
\sqrt{\pi}\,
\tau^{2k+\frac{3}{2}}
\frac{
\Gamma\!\left(\frac{\lambda}{2}+k+\frac{3}{2}\right)
\Gamma\!\left(\lambda+2k+\frac{3}{2}\right)
\Gamma\!\left(\frac{3}{2}\right)
}
{\Gamma\!\left(\frac{\lambda}{2}+k+2\right)}.
\end{equation}
We plot in figure \eqref{fig3}  how the Heisenberg uncertainty \eqref{Hei} varies with $n$ and $\tau$. Owing to the limited resolution of the plots,
 we numerically check the Heisenberg-like uncertainty \eqref{Hei} for some fixed values of the parameter $\tau$ in the Tables of the appendix \eqref{hu}. As results, one can see that for $\hbar=1$, $\tau=0.1$ and $\tau=0.5$ we have  $\sigma_x^2\sigma_p^2\geq  0.25$  for the all quantum levels except the ground state of the  system.

\section{Rényi and Tsallis entropies and  uncertainty based relations } \label{sec3}

Continuous Rényi and Tsallis entropies of the probability density $\rho_n^\lambda(z)$, with $(z=x,p)$ which characterizes the quantum  representation in position $\phi_n(x)$  and  in momentum $\phi_n(p)$  of a one-dimensional system, are defined, respectively, as
\begin{align}
\mathcal{R}^{(\alpha)}[\rho_n^{\lambda}(z)] &= 
\frac{1}{1-\alpha}\log\!\left(\mathcal{W}^{(\alpha)}[\rho_n^{\lambda}(z)]\right),
\qquad \alpha>0,\ \alpha\neq 1, 
\label{Re} \\
\mathcal{T}^{(\alpha)}[\rho_n^{\lambda}(z)] &= 
\frac{1}{1-\alpha}\log\left(\mathcal{W}^{(\alpha)}[\rho_n^{\lambda}(z)]-1\right),
\qquad \alpha>0,\ \alpha\neq 1. 
\label{Ts}
\end{align}
Here the quantity $\mathcal{W}^{(\alpha)}[\rho_n^{\lambda}(z)]$ denotes the entropic moment (or frequency moment) of order $\alpha$, defined by
\begin{equation}\label{wr}
\mathcal{W}^{(\alpha)}[\rho_n^{\lambda}(z)]
=\int_{\mathbb{R}} [\rho_n^\lambda(z)]^\alpha\,dz .
\end{equation}
It is worth noting that both continuous Rényi and Tsallis entropies reduce to the continuous Shannon entropy in the limit $\alpha \to 1$ \cite{7}. Indeed,
\begin{equation}\label{sha}
\mathcal{S}[\rho_n^\lambda(z)]
=
\lim_{\alpha\to1}\mathcal{R}^{(\alpha)}[\rho_n^\lambda(z)]
=
\lim_{\alpha\to1}\mathcal{T}^{(\alpha)}[\rho_n^\lambda(z)]
=
-\int_{\mathbb{R}}\rho_n^\lambda(z)\log\big(\rho_n^\lambda(z)\big)\,dz .
\end{equation}
Comparing to the  Heisenberg standard deviation,
these  quantities \eqref{sha} accurately and conveniently describe the extend or spread  of the position and momentum probability densities. Moreover, these entropies \eqref{sha} may discrease as well as increase, take negative or zero values. Therefore, increasing Shannon entropy   in the present context means more uncertainty on  the  particle position and  momentum probability densities. However, discreasing Shannon entropy means more predictability of particle probability densities. Negative and zero values of these entropies  indicate that probability densities are highly concentrated   and uniformly spread over a small region this space,respectively. Both Rényi and Tsallis information entropies exhibit similar letter features.

Given the relevance of entropic measures in characterizing quantum systems, we also consider the entropy-based uncertainty principle \cite{49f}.  Sobolev inequality of the conjugated Fourier transforms \cite{49g,49h}  is given by
\begin{eqnarray}\label{rt1}
    \left(\frac{\alpha}{\pi}\right)^{\frac{1}{4\alpha}}\left( \int_{\mathbb{R}}\rho_n^\alpha(x)dx\right)^{\frac{1}{2\alpha}}\geq \left(\frac{\beta}{\pi}\right)^{\frac{1}{4\beta}}\left( \int_{\mathbb{R}}\rho_n^\alpha(p)dp\right)^{\frac{1}{2\beta}},
\end{eqnarray}
with the non-negative coefficients $\alpha$ and $\beta$ obeying the constraint
\begin{eqnarray}\label{re}
    \frac{1}{\alpha}+\frac{1}{\beta}=2,\quad \mbox{with}\quad \frac{1}{2}<\alpha\leq 1.
\end{eqnarray}
Logarithmization of Equation \eqref{rt} yields the following inequality for the Rényi components
\begin{eqnarray}\label{rs}
    \mathcal{R}^{(\alpha)}[\rho_n^{\lambda}(x)]+\mathcal{R}^{(\beta)}[\rho_n^{\lambda}(p)] 
    &\geq&  -\frac{1}{2}\left(\frac{1}{1-\alpha}\log\frac{\alpha}{\pi}+\frac{1}{1-\beta}\log\frac{\beta}{\pi}\right), \quad   \mbox{with}\quad \frac{1}{\alpha}+\frac{1}{\beta}=2.
\end{eqnarray}
Within the limit of $\alpha\rightarrow 1$, equation \eqref{rs} degenerates to the Shannon uncertainty relation \cite{7,49g}:
\begin{eqnarray}
    \mathcal{S}[\rho_n^\lambda(x)]+ \mathcal{S}[\rho_n^\lambda(p)]=1+\log \pi.
\end{eqnarray}

By rewriting the Sobolev inequality \eqref{rt} in terms of the Tsallis entropy formulation \eqref{Ts} with the same constraints \eqref{re}, the Tsallis entropy-based uncertainty relation \cite{43} is obtained.
\begin{eqnarray}\label{rp}
    \left(\frac{\alpha}{\pi}\right)^{\frac{1}{4\alpha}}\left( (1-\alpha)\mathcal{T}^{(\alpha)}[\rho_n^{\lambda}(x)]+1\right)^{\frac{1}{2\alpha}}\geq \left(\frac{\beta}{\pi}\right)^{\frac{1}{4\beta}}\left( (1-\beta)\mathcal{T}^{(\alpha)}[\rho_n^{\lambda}(p)]+1\right)^{\frac{1}{2\beta}}.
\end{eqnarray}

With the latter equations at hand,  we compute Rényi and Tsallis entropies in position  and momentum representations and the corresponding entropy-like based uncertainty relations.

\subsection{Rényi and Tsallis entropies in position representation}
Rényi and Tsallis entropies in position representations, are obtained  by computing the  entropic moment \eqref{wr} given by
\begin{equation}\label{as}
\mathcal{W}^{(\alpha)}[\rho_n^{\lambda}(x)]=
\int_{\mathbb{R}}[\rho_n^\lambda(x)]^\alpha dx .
\end{equation}
Subtituting the probability density function \eqref{p1} in equation \eqref{as}, we have
\begin{equation}\label{as1}
\mathcal{W}^{(\alpha)}[\rho_n^{\lambda}(x)]=
N_n^{2\alpha}
\int_{-\infty}^{+\infty}
(1+\tau x^2)^{-\alpha(\lambda+1)}
\left|C_n^\lambda\!\left(\frac{x\sqrt{\tau}}{\sqrt{1+\tau x^2}}\right)\right|^{2\alpha}
dx .
\end{equation}
Using the change of variable \eqref{wt}, equation \eqref{as} is reduced into   
\begin{eqnarray}\label{pa2}
\mathcal{W}^{(\alpha)}[\rho_n^{\lambda}(x)]=
\frac{N_n^{2\alpha}}{\sqrt{\tau}}
\int_{-1}^{+1}
(1-y^2)^{\alpha(\lambda+1)-\frac{3}{2}}
\left[C_n^\lambda(y)\right]^{2\alpha}dy.
\end{eqnarray}

The functional integral \eqref{pa2}  is a particular case of  the Rényi-like functional of the Gegenbauer polynomials \cite{48,49,49a,49b,49c,49d}. These functional integrals are solved for large positive values of the parameter $\lambda$ \cite{49a,49b,49c,49d}. So, around  the  asymptotic $(\tau\rightarrow 0 \implies\lambda \rightarrow \infty)$ which corresponds to the undeformed harmonic oscillator, we have the limiting relation
\begin{eqnarray}\label{bii}
    \lim_{\lambda\rightarrow \infty} \frac{C_n^{(\lambda)}(y)}{C_n^{(\lambda)}(1)}=y^n\quad \mbox{with}\quad C_n^{(\lambda)}(1)=\frac{\Gamma(n+2\lambda)}{\Gamma(n+1) \Gamma(2\lambda)}.
    \end{eqnarray}
Using this limiting relation \eqref{bii} in Equation \eqref{pa2}, obtaining the asymptotics
\begin{eqnarray}\label{pa2}
\mathcal{W}^{(\alpha)}[\rho_n^{\lambda}(x)]\sim \frac{[N_n C_n^{(\lambda)}(1)]^{2\alpha}}{\sqrt{\tau}}
\int_{-1}^{+1} y^{2\alpha n}
(1-y^2)^{\alpha(\lambda+1)-\frac{3}{2}}
dy.
\end{eqnarray}
Using the Beta function identity \cite{55}
\begin{equation}\label{wr}
\int_{-1}^{1} y^{2m}(1-y^2)^{\xi}dy
=
B\!\left(m+\frac12,\xi+1\right)= \frac{\Gamma(m+\frac{1}{2})\Gamma(\xi+1)}{\Gamma(m+\xi+\frac{3}{2})},
\end{equation}
and by corresponding $\xi=\alpha(\lambda+1)-3/2$ and $m= n\alpha$,   equation \eqref{pa2} is reduced into (See appendix\eqref{xv1}) 
\begin{eqnarray}\label{rt}
\mathcal{W}^{(\alpha)}[\rho_n^{\lambda}(x)]&\sim& \frac{[N_n C_n^{(\lambda)}(1)]^{2\alpha}}{\sqrt{\tau}}
\frac{\Gamma\!\left(n\alpha+\frac12\right)
\Gamma\!\left(\alpha(\lambda+1)-\frac12\right)}
{\Gamma\!\left(n\alpha+\alpha(\lambda+1)\right)}\cr
&\sim&
\frac{1}{\sqrt{\tau}}
\frac{
2^{n\alpha}
\Gamma\!\left(n\alpha+\frac12\right)
}{
(n!)^\alpha
\pi^{\alpha/2}
\alpha^{\,n\alpha+\frac12}
}
\lambda^{\frac{\alpha-1}{2}},
\qquad
\lambda\to\infty. \label{zza}
\end{eqnarray}
 Note that this expression of the entropy moment for large positive value $\lambda$ represents the entropy moment of an ordinary harmonic oscillator. This can be explicitly shown  by computing  some particular cases of  equation \eqref{rt} for some values of the states $n$:
\begin{itemize}
    \item For the ground state  $(n = 0)$,  
     we get the  entropy moment of  the ground state of undeformed harmonic oscillator similar to the one recently obtained in \cite{43}
    \begin{eqnarray}\label{wa}
\mathcal{W}^{(\alpha)}[\rho_0^{\lambda}(x)]
 \sim
\left(\frac{\lambda}{\pi}\right)^{\frac{\alpha-1}{2}}\frac{1}{\sqrt{\tau\alpha}},
\qquad \lambda\to\infty.
\end{eqnarray}
 As we can see in \cite{43}, the ground state of the undeformed  harmonic oscillator is independent of the deformed parameter. However, in the present context we can  see  that equation \eqref{wa} is simultaneously dependent of the deformed parameters $\lambda$ and $\tau$. This is due to  the  asymptotic behavior of this model.
\item For the first excited state  $(n= 1)$,  we have 
 \begin{eqnarray}
\mathcal{W}^{(\alpha)}[\rho_1^{\lambda}(x)]
\sim
\frac{1}{\sqrt{\tau}}\frac{2^\alpha \Gamma(\alpha +\frac{1}{2})}{\pi^{\alpha/2}\alpha^{\alpha+\frac{1}{2}}}\lambda^{\frac{\alpha-1}{2}} \qquad \lambda\to\infty.
\end{eqnarray}

\end{itemize}

The Rényi entropy \eqref{Re} in position representation  becomes
\begin{eqnarray}\label{rst}
\mathcal{R}^{(\alpha)}[\rho_n^{\lambda}(x)] 
\sim
\frac{1}{1-\alpha} 
\log \Biggl[ \frac{1}{\sqrt{\tau}}
\frac{
2^{n\alpha}
\Gamma\!\left(n\alpha+\frac12\right)
}{
(n!)^\alpha
\pi^{\alpha/2}
\alpha^{\,n\alpha+\frac12}
}
\lambda^{\frac{\alpha-1}{2}}
\Biggr], \qquad \lambda\to\infty.
\end{eqnarray}
Remarkably,  we obtain the Rényi entropy for  the ground, $n = 0$, and first excited, $n = 1,$ states recently obtained in \cite{43,49e} for the undeformed harmonic oscillator
\begin{eqnarray}
\mathcal{R}^{(\alpha)}[\rho_0^{\lambda}(x)] 
&\sim&
\frac{1}{1-\alpha} 
\log \Biggl[ \left(\frac{\lambda}{\pi}\right)^{\frac{\alpha-1}{2}}\frac{1}{\sqrt{\tau\alpha}}
\Biggr], \qquad \lambda\to\infty\cr
&\sim&  \frac{1}{2} \log\left(\frac{\pi}{\lambda}\right)-\frac{1}{1-\alpha}\log\sqrt{\tau\alpha}, \qquad \lambda\to\infty.\\
\mathcal{R}^{(\alpha)}[\rho_1^{\lambda}(x)] 
&\sim&
\frac{1}{1-\alpha} 
\log \Biggl[ \frac{1}{\sqrt{\tau}}\frac{2^\alpha \Gamma(\alpha +\frac{1}{2})}{\pi^{\alpha/2}\alpha^{\alpha+\frac{1}{2}}}\lambda^{\frac{\alpha-1}{2}}
\Biggr], \qquad \lambda\to\infty\cr
&\sim& -\frac{1}{2}\log\lambda +\frac{1}{1-\alpha} 
\log \Biggl[\frac{2^\alpha }{\tau^{1/2}\pi^{\alpha/2}\alpha^{\alpha+1/2 }} \Gamma\left(\alpha +\frac{1}{2}\right)
\Biggr] \qquad \lambda\to\infty.
\end{eqnarray}

Similarly, the Tsallis entropy \eqref{Ts} in position representation reads
\begin{eqnarray} \label{Tsr1}
\mathcal{T}^{(\alpha)}[\rho_n^{\lambda}(x)] 
\sim
\frac{1}{1-\alpha} \log
\Biggl[  \frac{1}{\sqrt{\tau}}
\frac{
2^{n\alpha}
\Gamma\!\left(n\alpha+\frac12\right)
}{
(n!)^\alpha
\pi^{\alpha/2}
\alpha^{\,n\alpha+\frac12}
}
\lambda^{\frac{\alpha-1}{2}}
-1\Biggr], \qquad \lambda\to\infty.
\end{eqnarray}
 The Tsallis entropy  for the ground state $n = 0$ and for the first excited state $n = 1,$ are analytically similar to the ones obtained in \cite{43,49e}
 \begin{eqnarray}
\mathcal{T}^{(\alpha)}[\rho_0^{\lambda}(x)] 
&\sim&
\frac{1}{1-\alpha} 
\log\Biggl[\left(\frac{\lambda}{\pi}\right)^{\frac{\alpha-1}{2}}\frac{1}{\sqrt{\tau\alpha}} -1 \Biggr], \qquad \lambda\to\infty.\\
\mathcal{T}^{(\alpha)}[\rho_1^{\lambda}(x)]
&\sim&
\frac{1}{1-\alpha} 
\Biggl[ \frac{1}{\sqrt{\tau}}\frac{2^\alpha \Gamma(\alpha +\frac{1}{2})}{\pi^{\alpha/2}\alpha^{\alpha+\frac{1}{2}}}\lambda^{\frac{\alpha-1}{2}}-1 \Biggr], \qquad \lambda\to\infty.
\end{eqnarray}

\begin{figure}
\centering
\includegraphics[width=7.5cm, height=8cm]{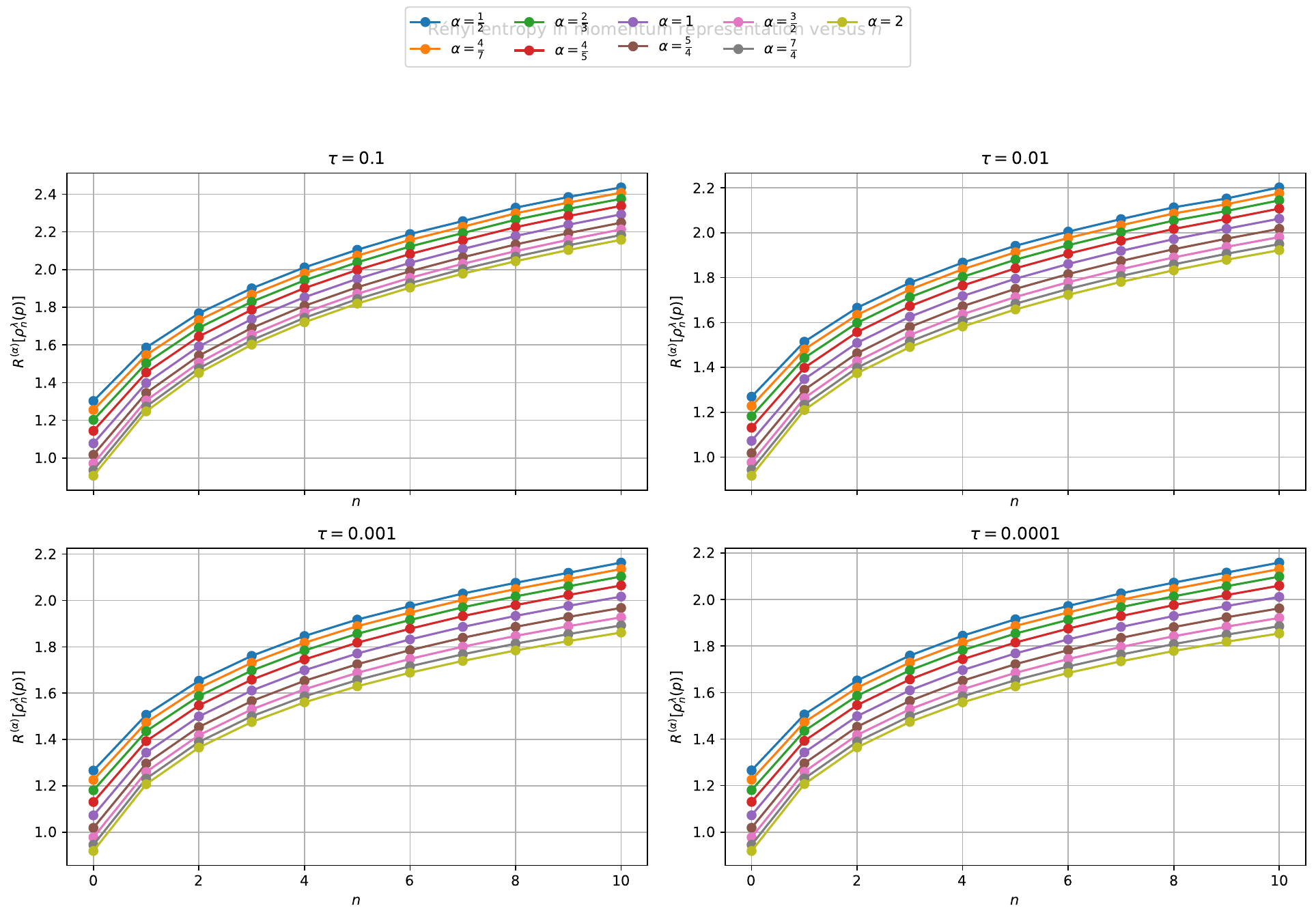}
\includegraphics[width=7.5cm, height=8cm]{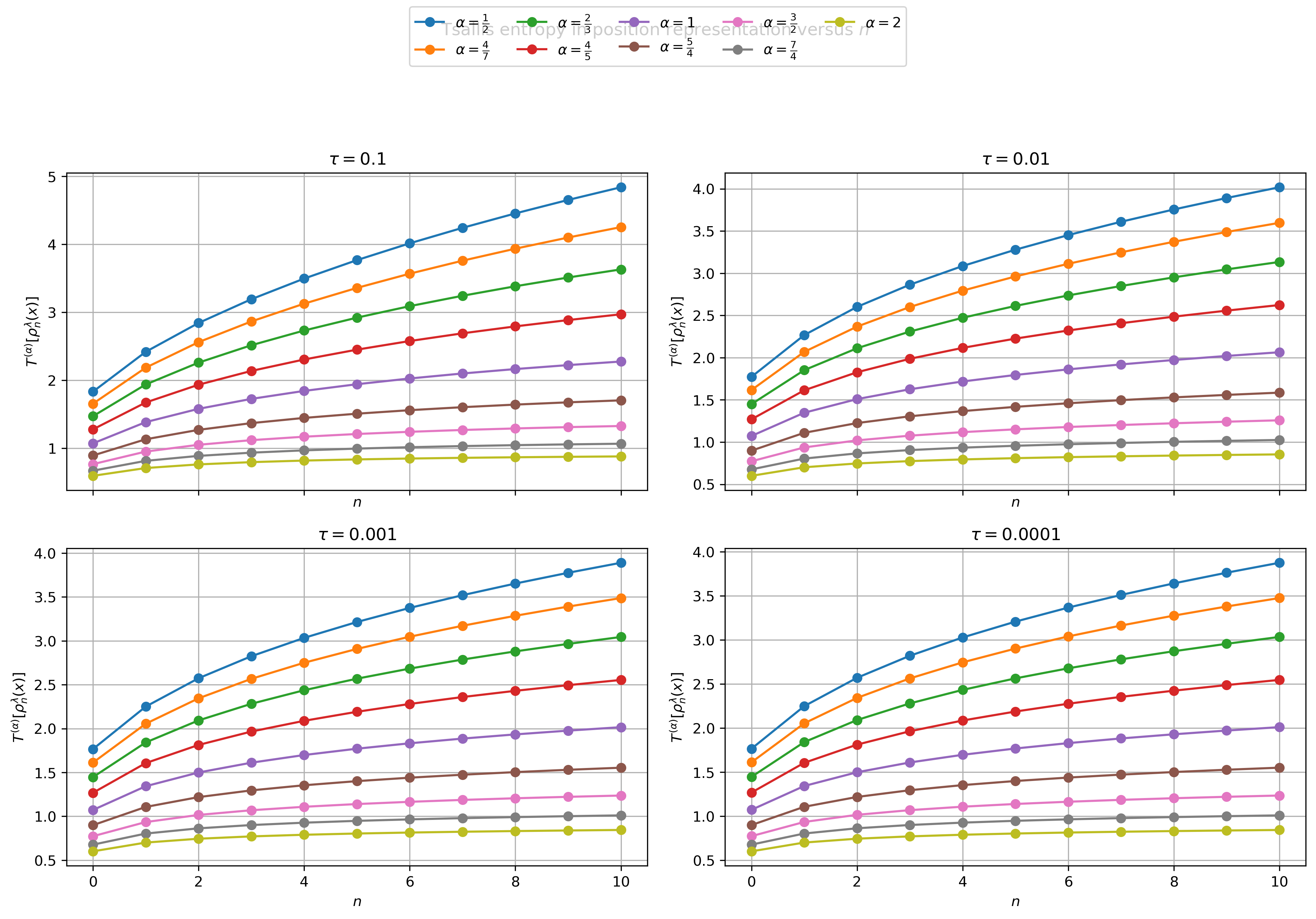}
\caption{Rényi entropy \eqref{rst} (left 4-subplots) and Tsallis entropy\eqref{Tsr1} (right 4-subplots) as functions $n$ for fixed values of parameters $\alpha $ and $\tau$. }.
\label{fg411}
\end{figure}

In   figure \eqref{fg411},  we analyse the  behavior  of 
Rényi and Tsallis entropies at the asymptotic limit  $(\tau\rightarrow 0 \implies \lambda\rightarrow \infty)$ of the system which  that depicts the behavior of an undeformed harmonic oscillator  with respect to the
parameters $n$ and $\alpha$.  It also depicts the behavior both below and above the Shannon
case ($\alpha \rightarrow 1$). As we can see, Rényi and Tsallis entropies approach the Shannon entropy when $\alpha \rightarrow 1$.
Both Rényi and Tsallis entropies monotonically increase with the parameters $n$ and $\alpha$, but in case of Tsallis entropy, the curves become progressively closer for higher values of $n$. These observation are similar to the one recently obtained in \cite{43}.
.

\subsection{Rényi and Tsallis entropies in momentum representation}

\begin{figure}
\centering
\includegraphics[width=7cm, height=8cm]{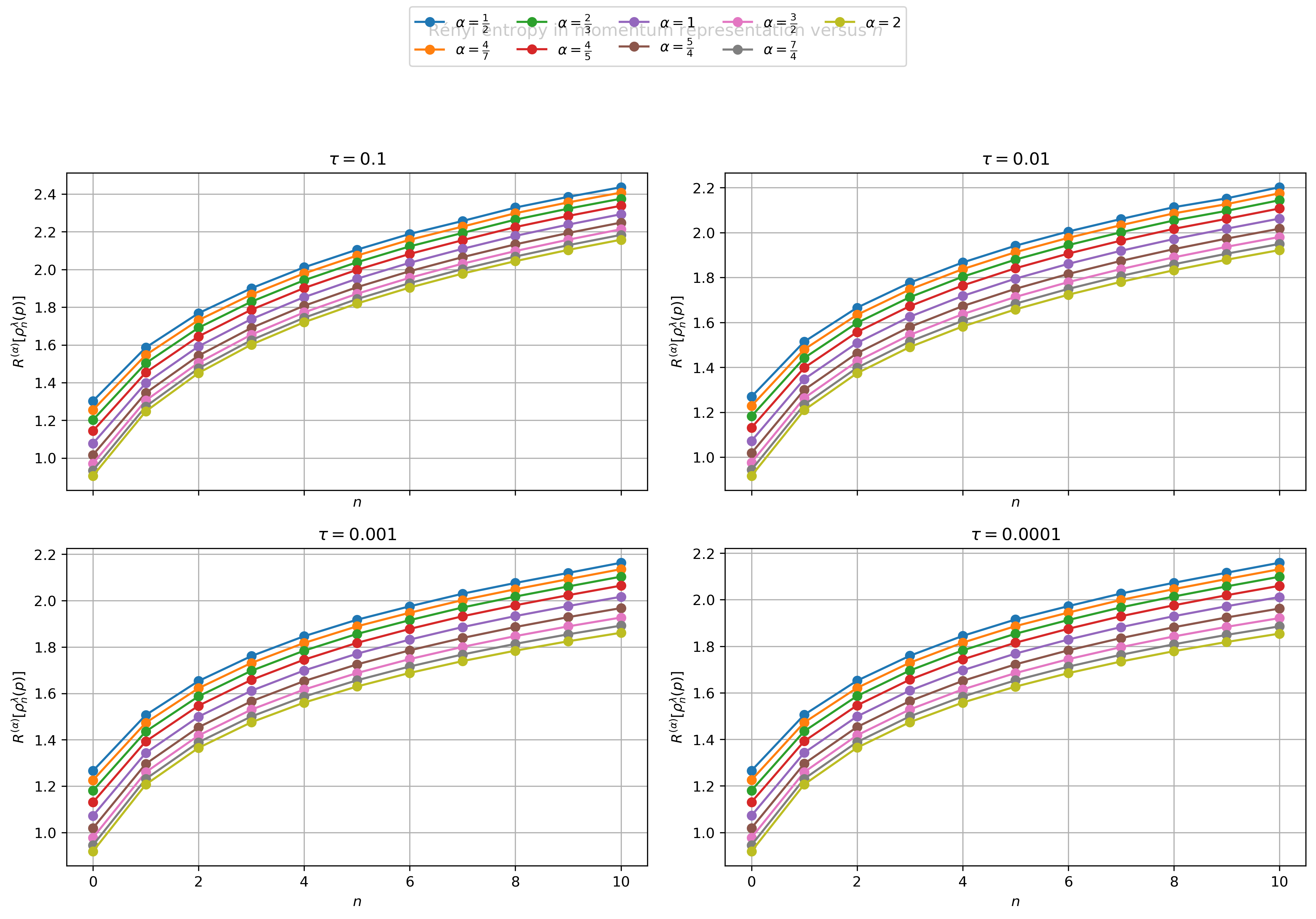}
\includegraphics[width=7cm, height=8cm]{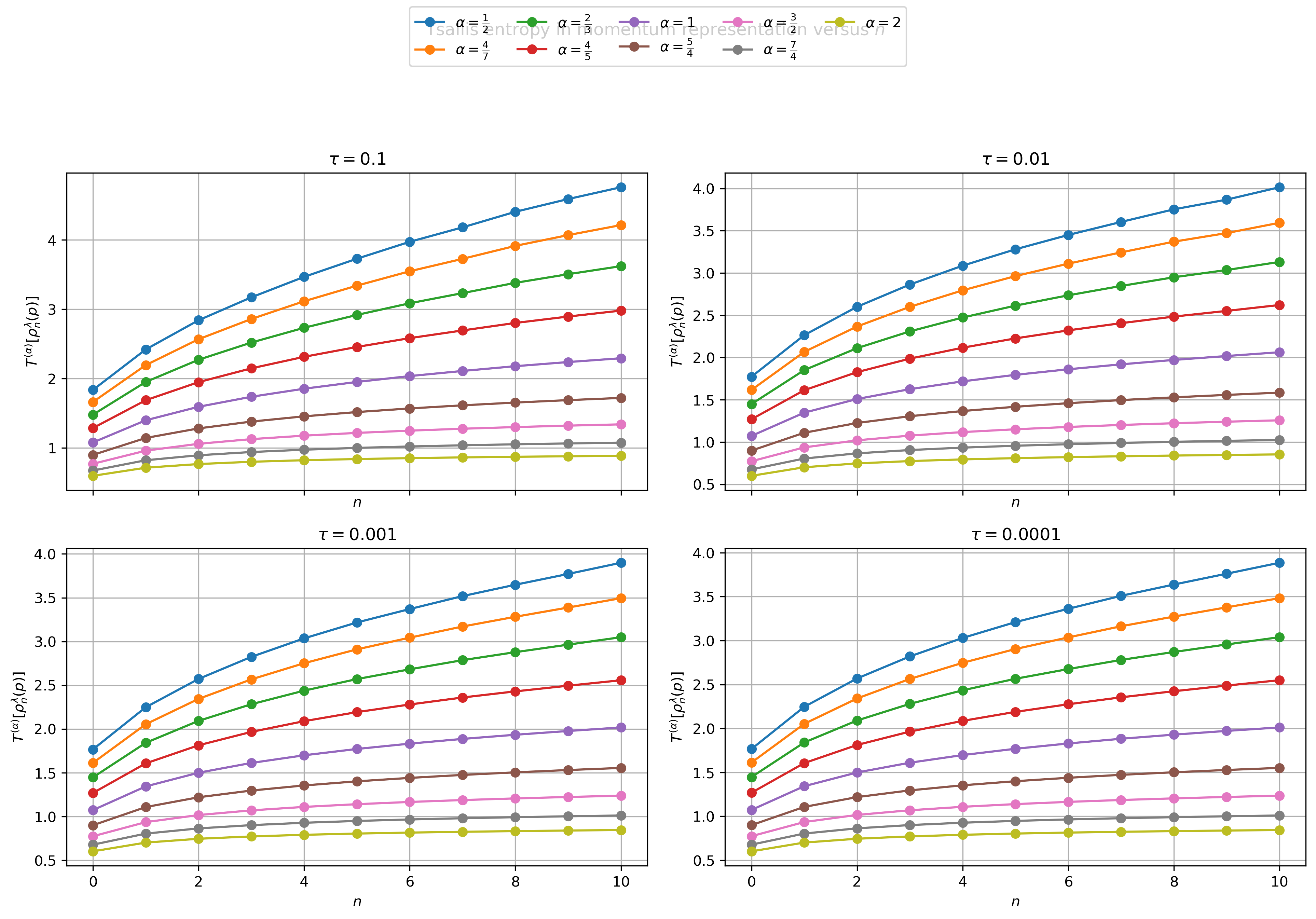}
\caption{Rényi entropy \eqref{renq} (left 4-subplots) and Tsallis entropy \eqref{qrt} (right 4-subplots) in momentum	representations as functions $n$ for fixed values of parameters $\alpha$ and $\tau$ }.
\label{fg5y}
\end{figure}

The entropic moment of order $\alpha$ in momentum representation is given as
\begin{equation}
 \mathcal{W}^{(\alpha)}[\rho_n^\lambda(p)]
=
\int_{\mathbb{R}} \left[\rho_n^{\lambda}(p)\right]^{\alpha}\,dp,
\qquad \alpha > 0,\ \alpha \neq 1.
\end{equation}
Now we substitute the explicit expression of $\rho_n^{\lambda}(p)$, and obtain
\begin{equation}\label{xv}
\mathcal{W}^{(\alpha)}[\rho_n^\lambda(p)]
= \left[A_{n,k}^2
\right]^{\alpha} \int_{-\infty}^{+\infty}
\left(\frac{|p|}{2\sqrt{\tau}}\right)^{\lambda\alpha+2\alpha k}
K_{\frac{\lambda}{2}+k}^{\,2\alpha}\!\left(\frac{|p|}{\sqrt{\tau}}\right)\,dp.
\end{equation}
Using the change of variable \eqref{tr}, the entropic moment \eqref{xv} is given by
\begin{equation}
\mathcal{W}^{(\alpha)}[\rho_n^\lambda(p)] = \left[A_{n,k}^2
\right]^{\alpha}2^{1-(\lambda\alpha+2\alpha k)}\sqrt{\tau}\int_0^{+\infty}
u^{\lambda \alpha+2\alpha k}K_{\frac{\lambda}{2}+k}^{\,2\alpha}(u)\,du.
\end{equation}
The last integral can be evaluated as follows
\begin{equation}
\int_0^{+\infty} u^{2\alpha k+\lambda\alpha}
K_{\frac{\lambda}{2}+k}^{\,2\alpha}(u)\,du = \frac{2^{2 k + \lambda - 2} \sqrt{\pi} \Gamma^{2}\left(k + \frac{\lambda}{2} + \frac{1}{2}\right) \Gamma\left(2 k + \lambda + \frac{1}{2}\right)}{\Gamma\left(2 k + \lambda + 1\right)}.
\end{equation}
Finally, we have
\begin{equation}\label{rtw1}
\mathcal{W}^{(\alpha)}[\rho_n^\lambda(p)] = \left[A_{n,k}^2
\right]^{\alpha}
2^{-(\lambda\alpha+2\alpha k)+2k+\lambda-1}\sqrt{\pi\tau} \frac{\Gamma^{2}\left(k + \frac{\lambda}{2} + \frac{1}{2}\right) \Gamma\left(2 k + \lambda + \frac{1}{2}\right)}{\Gamma\left(2 k + \lambda + 1\right)}.
\end{equation}
Analyzing certain specific instances of equation \eqref{rtw1} in cas $\lambda\rightarrow \infty$ (harmonic oscillator)  and  with some values of the states $n$ $(n=0,\, n=1)$ are  obtained as follows:
\begin{itemize}
\item For the ground state $n=0$, the equation \eqref{rtw1} is reduced into 
\begin{equation}\label{rtw}
\mathcal{W}^{(\alpha)}[\rho_0^\lambda(p)] = \left[A_{0,k}^{2}\right]^{\alpha}
2^{\,2k+\lambda-1-\alpha(\lambda+2k)}
\sqrt{\pi\tau}\,
\frac{
\Gamma^{2}\!\left(k+\frac{\lambda+1}{2}\right)
\Gamma\!\left(2k+\lambda+\frac12\right)
}
{\Gamma\!\left(2k+\lambda+1\right)},
\end{equation}
with the normalized constant $\left[A_{0,k}^{2}\right]^{\alpha}$ given by
\begin{equation}
A_{0,k}^{2}
=
\frac{
\Gamma(2k+\lambda+1)
}
{
\Gamma^{2}\!\left(k+\frac{\lambda+1}{2}\right)
}.
\end{equation}

\item For the first excited state $(n = 1)$, we have
 \begin{equation}\label{rtw}
\mathcal{W}^{(\alpha)}[\rho_1^\lambda(p)]
=
\left[A_{1,k}^2\right]^{\alpha}
2^{\,2k+\lambda-1-\alpha(\lambda+2k)}
\sqrt{\pi\tau}\,
\frac{
\Gamma^{2}\!\left(k+\frac{\lambda+1}{2}\right)
\Gamma\!\left(2k+\lambda+\frac12\right)
}
{
\Gamma\!\left(2k+\lambda+1\right)
},
\end{equation}
where the normalized constant $A_{1,k}^{2}$ is given by
\begin{equation}
A_{1,k}^{2}
=
\frac{\Gamma(2k+\lambda+2)}
{\Gamma^{2}\!\left(k+\frac{\lambda+3}{2}\right)}.
\end{equation}

    \item For large positive values of the parameter 
    $\lambda$ which corresponds to the ordinary harmonic oscillator, the equation \eqref{rtw}
    \begin{equation}\label{rtw3}
 \mathcal{W}^{(\alpha)}[\rho_n^\infty(p)] \sim \left[A_{n,k}^{2}\right]^{\alpha}
\pi\sqrt{\pi\tau}\,
2^{2k(1-\alpha)}
\lambda^{2k-\frac12}
\left(\frac{\lambda}{2^{\alpha}e}\right)^{\lambda},
\qquad \lambda\to\infty.
\end{equation}

The Rényi entropy \eqref{Re} in momentum representation  becomes
\begin{eqnarray}\label{renq}
\mathcal{R}^{(\alpha)}[\rho_n^{\lambda}(p)] 
=
\frac{1}{1-\alpha} 
\log \Biggl[ \left[A_{n,k}^2
\right]^{\alpha} 2^{-(\lambda\alpha+2\alpha k)+2k+\lambda-1} \sqrt{\pi\tau} \frac{\Gamma^{2}\left(k + \frac{\lambda}{2} + \frac{1}{2}\right) \Gamma\left(2 k + \lambda + \frac{1}{2}\right)}{\Gamma\left(2 k + \lambda + 1\right)} 
\Biggr] .
\end{eqnarray}
The Rényi entropy for  some special cases are given by
\begin{eqnarray}
\mathcal{R}^{(\alpha)}[\rho_0^{\lambda}(p)] 
&=&
\frac{1}{1-\alpha} 
\log \Biggl[
\left[A_{0,k}^2
\right]^{\alpha}2^{\,2k+\lambda-1-\alpha(\lambda+2k)}
\sqrt{\pi\tau}\,
\frac{
\Gamma^{2}\!\left(k+\frac{\lambda+1}{2}\right)
\Gamma\!\left(2k+\lambda+\frac12\right)
}
{\Gamma\!\left(2k+\lambda+1\right)} \Biggr].\\
\mathcal{R}^{(\alpha)}[\rho_1^{\lambda}(p)] 
&=&
\frac{1}{1-\alpha} 
\log \Biggl[
\left[A_{1,k}^2
\right]^{\alpha}2^{\,2k+\lambda-1-\alpha(\lambda+2k)}
\sqrt{\pi\tau}\,
\frac{
\Gamma^{2}\!\left(k+\frac{\lambda+1}{2}\right)
\Gamma\!\left(2k+\lambda+\frac12\right)
}
{\Gamma\!\left(2k+\lambda+1\right)} \Biggr].\\
\mathcal{R}^{(\alpha)}[\rho_n^{\infty}(p)] &=& \frac{1}{1-\alpha} 
\log \Biggl[ \left[A_{n,k}^{2}\right]^{\alpha}
\pi\sqrt{\pi\tau}\,
2^{2k(1-\alpha)}
\lambda^{2k-\frac12}
\left(\frac{\lambda}{2^{\alpha}e}\right)^{\lambda} \Biggr],
\qquad \lambda\to\infty.
\end{eqnarray}

Similarly, the Tsallis entropy \eqref{Ts} in momentum representation reads
\begin{eqnarray}\label{qrt}
\mathcal{T}^{(\alpha)}[\rho_n^{\lambda}(p)] 
=
\frac{1}{1-\alpha} \log
\Biggl[ \left[A_{n,k}^2
\right]^{\alpha} 2^{-(\lambda\alpha+2\alpha k)+2k+\lambda-1} \sqrt{\pi\tau} \frac{\Gamma^{2}\left(k + \frac{\lambda}{2} + \frac{1}{2}\right) \Gamma\left(2 k + \lambda + \frac{1}{2}\right)}{\Gamma\left(2 k + \lambda + 1\right)}
-1\Biggr] .
\end{eqnarray}
The  Tsallis entropy for  some special cases are given by
\begin{eqnarray}
\mathcal{T}^{(\alpha)}[\rho_0^{\lambda}(p)] 
&=&
\frac{1}{1-\alpha}\times\cr&& \log
\Biggl[ \left[A_{0,k}^2
\right]^{\alpha} 2^{-(\lambda\alpha+2\alpha k)+2k+\lambda-1} \sqrt{\pi\tau} \frac{\Gamma^{2}\left(k + \frac{\lambda}{2} + \frac{1}{2}\right) \Gamma\left(2 k + \lambda + \frac{1}{2}\right)}{\Gamma\left(2 k + \lambda + 1\right)}
-1\Biggr].\\
\mathcal{T}^{(\alpha)}[\rho_1^{\lambda}(p)] 
&=&
\frac{1}{1-\alpha}\times\cr&& \log
\Biggl[ \left[A_{1,k}^2
\right]^{\alpha} 2^{-(\lambda\alpha+2\alpha k)+2k+\lambda-1} \sqrt{\pi\tau} \frac{\Gamma^{2}\left(k + \frac{\lambda}{2} + \frac{1}{2}\right) \Gamma\left(2 k + \lambda + \frac{1}{2}\right)}{\Gamma\left(2 k + \lambda + 1\right)}
-1\Biggr].\\
\mathcal{T}^{(\alpha)}[\rho_n^{\infty}(p)] 
&=&
\frac{1}{1-\alpha} \log
\Biggl[ \left[A_{n,k}^{2}\right]^{\alpha}
\pi\sqrt{\pi\tau}\,
2^{2k(1-\alpha)}
\lambda^{2k-\frac12}
\left(\frac{\lambda}{2^{\alpha}e}\right)^{\lambda}-1 \Biggr],
\qquad \lambda\to\infty.
\end{eqnarray}
 Figure\eqref{fg5y} explores  the  behavior of continuous Rényi \eqref{renq} and Tsallis \eqref{qrt}  entropies in momentum representation as function of the quantum $n$ for fixed values of a parameter $\alpha$. As expected, both
 entropies converge to the Shannon entropy at the limit $\alpha\rightarrow 1$. We also see that, at the asymptotic limit $(\tau\rightarrow 0 \implies \lambda\rightarrow\infty)$ which depicts the behavior of an ordinary harmonic oscillator,  both entropies in momentum space closely resembles that in position space from figure \eqref{fg411}.

 \end{itemize}

\subsection{Rényi and Tsallis information uncertainty relations}
\begin{figure}
\centering
\includegraphics[width=6.5cm, height=6cm]{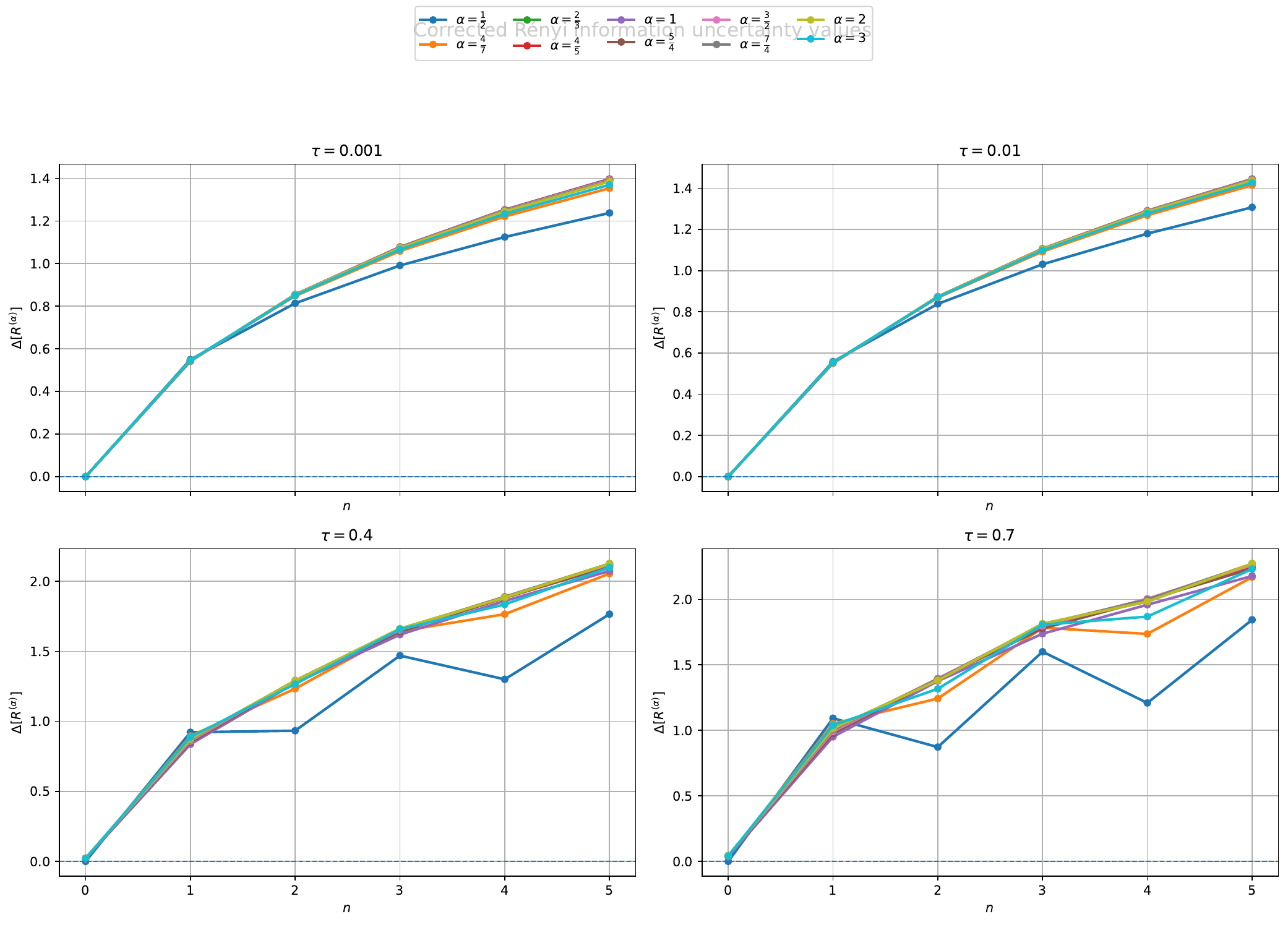}
\includegraphics[width=6.5cm, height=6cm]{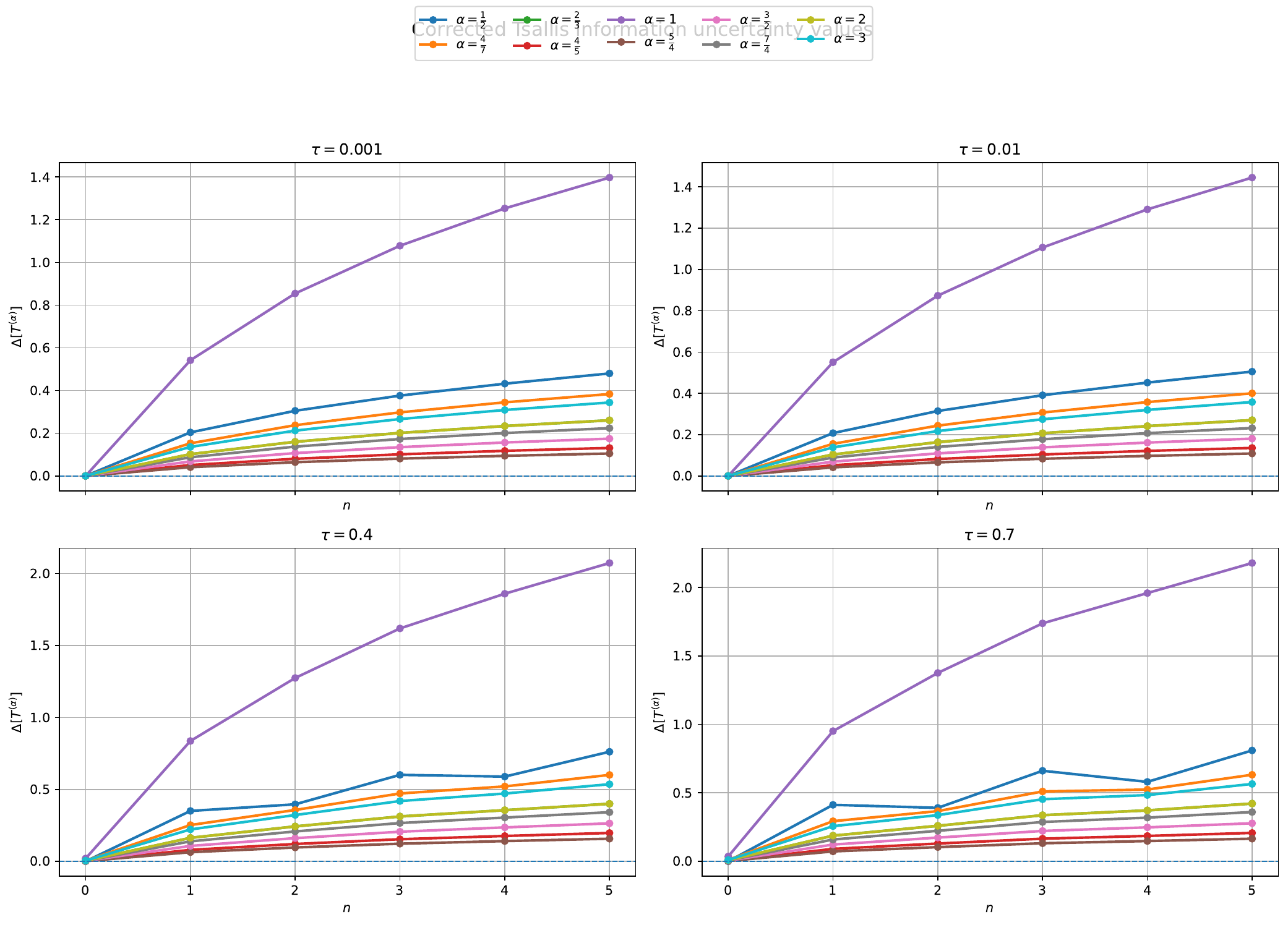} 
\caption{ Rényi  uncertainty entropy  \eqref{wz1} (left 4-subplots)  and Rényi uncertainty entropy \eqref{wz2}  (right 4-subplots) as functions $n$ for fixed values of parameters $\alpha$ and $\tau$}.
\label{fg6}
\end{figure}

Rényi entropy and Tsallis information entropies quantify the average uncertainty by weighting  probability distributions differently than Shannon  entropy. In the present context, we consider a given entropy functional $X$ and $\Delta$ the associated uncertainty function  such that $\Delta [X]\geq 0$
and the equality is reached only when the uncertainty relation is saturated. Based on the latter, for  the system at hand, the uncertainty $\Delta$ associates to the  difference  between the two sides of the  Rényi\eqref{rs} and Tsallis\eqref{rp} uncertainty  entropies reads, respectively
\begin{eqnarray}
   \Delta\left[\mathcal{R}^{(\alpha)} \right]&=&\mathcal{R}^{(\alpha)}[\rho_n^{\lambda}(x)]+\mathcal{R}^{(\beta)}[\rho_n^{\lambda}(p)] 
    + \frac{1}{2}\left(\frac{1}{1-\alpha}\log\frac{\alpha}{\pi}+\frac{1}{1-\beta}\log\frac{\beta}{\pi}\right), \label{wz1}  \\
    \Delta\left[\mathcal{T}^{(\alpha)}  \right]&=&  \left(\frac{\alpha}{\pi}\right)^{\frac{1}{4\alpha}}\left( (1-\alpha)\mathcal{T}^{(\alpha)}[\rho_n^{\lambda}(x)]+1\right)^{\frac{1}{2\alpha}}- \left(\frac{\beta}{\pi}\right)^{\frac{1}{4\beta}}\left( (1-\beta)\mathcal{T}^{(\beta)}[\rho_n^{\lambda}(p)]+1\right)^{\frac{1}{2\beta}}, \label{wz2}
\end{eqnarray}
where $ \frac{1}{\alpha}+\frac{1}{\beta}=2$ or $\beta= \frac{\alpha}{2\alpha-1}$ in both expressions. 
We plot in figure\eqref{fg6} how entropy uncertainty functions $ \Delta\left[\mathcal{R}^{(\alpha)} \right]$ and $ \Delta\left[\mathcal{T}^{(\alpha)}  \right]$ varies with $n$, $\alpha$ and $\tau$. One can observe that, the ground state
saturates both entropy uncertainty functions as in \cite{43} and therefore the uncertainty function vanishes. This previous point is no longer true from the first excited states. So, both uncertainty functions
increase with the quantum number $n$ but decrease with the parameter $\alpha$ meanning that both Rényi and Tsallis entropies are more sensitive to variations of $\alpha$ in the momentum space than in position
space. This behaviour is consistent with the one recently obtained in \cite{43}. Owing to the limited
resolution of the plots, this effect can be more clearly appreciated in the Tables \eqref{tab2} and \eqref{tab} of the appendix \eqref{vb}.

\section{Conclusion}\label{sec4}
In this paper, we studied  continuous  Rényi and Tsallis information entropies for a  position dependent mass confined in a harmonic oscillator. This study  provided a deeper understanding of how these information entropies measure how particles of this model  are  distributed in position and momentum representations. We finally evaluated   the related entropic uncertainty relations numerically to validate the the latter observations.

The result achieved in this paper has been  gone through some   mathematical techniques and physical interpretations. In fact, the position  and the momentum probability distributions of the model are controlled by Gegenbauer polynomials  and the modified Bessel
function of the second kind respectively. We have shown that in both representations, the particles of the system are highly distributed in a small region around the origin. To confirm the latter observation, we have studied Rényi and Tsallis information entropies associated to these  distributions.  In fact these entropies  generalize the Shannon entropy  and provide a tunable parameter that highlights different aspects of a probability distribution. Because the Renyi and Tsallis information
entropies in position representation are represented by integral functionals of the Gegenbauer polynomials, these quantities are much more challenging to calculate. We have therefore  wanalytically and numerically evaluated the asymptotical
behaviors of these entropies using some modern techniques of approximation theory. The results are similar to those recently obtained  for the undeformed harmonic oscillator \cite{43,49e}.  However, the Tsallis and R´enyi information entropies
in momentum representation have been obtained without the need of any approximation technique.
At the asymptotic limit $(\tau\rightarrow 0 \implies \lambda \rightarrow\infty)$ which depicts the behavior of an ordinary harmonic
oscillator, both entropies in momentum space closely resembles that in position space and are closed to
the results of similar models of the literature \cite{43}. Finally, we evaluated the related entropic uncertainty
relations numerically to validate the latter observations.


\section*{Appendix A: Determination of equation \texorpdfstring{\eqref{qa}}{(87)}}\label{xv}

Indeed, using the following relations~\cite{55}
\begin{equation}
	\lim_{\lambda\to\infty}
	\lambda^{-n/2}
	C_n^{(\lambda)}
	\left(\frac{x}{\sqrt{\lambda}}\right)
	=
	\frac{1}{n!}H_n(x),
	\qquad
	\lim_{\lambda\to\infty}
	\frac{\Gamma(\lambda+a)}{\Gamma(\lambda)}
	e^{-a\ln\lambda}
	=1,
	\label{eq:84}
\end{equation}

the doubling formula
\begin{equation}
	\Gamma(2x)
	=
	\frac{2^{2x-1}}{\sqrt{\pi}}
	\Gamma(x)
	\Gamma\left(x+\frac{1}{2}\right),
	\label{eq:85}
\end{equation}

and the expansion
\[
\ln(1+\tau x^2)\sim\tau x^2,
\qquad \tau\to0,
\]
we obtain
\begin{equation}
	\left(1+\tau x^2\right)^{-\frac{\lambda+\frac12}{2}}
	=
	\exp\left[
	-\frac{\lambda+\frac12}{2}
	\ln(1+\tau x^2)
	\right]
	\sim
	\exp\left(-\frac{\kappa x^2}{2}\right),
	\label{eq:86}
\end{equation}
where
\begin{equation}
	\kappa=\lambda\tau=\frac{m_0\omega}{\hbar}.
\end{equation}

Combining the above results, we find that the wavefunctions converge
\begin{equation}
	\lim_{\tau\to0}\phi_n^\tau(x)
	=
	\phi_n^\infty(x)
	=
	\frac{1}{\sqrt{2^n n!}}
	\left(\frac{m_0\omega}{\pi\hbar}\right)^{1/4}
	e^{-\frac{m_0\omega}{2\hbar}x^2}
	H_n\left(
	\sqrt{\frac{m_0\omega}{\hbar}}\,x
	\right).
	\label{eq:87}
\end{equation}
\section*{Appendix B: Proof of equation  \texorpdfstring{\eqref{zza}}{}}\label{xv1}

In this subsection, we show how the asymptotic behaviour of the entropic moment (55) is obtained for $\lambda \to \infty$ which corresponds $\tau \to 0$. Starting from the following equation,
\begin{equation}
	\mathcal{W}^{(\alpha)}[\rho_n^\lambda(x)] \sim \frac{[N_n C_n^{(\lambda)}(1)]^{2\alpha}}{\sqrt{\tau}} \frac{\Gamma\left(n\alpha + \frac{1}{2}\right)\Gamma\left(\alpha(\lambda+1) - \frac{1}{2}\right)}{\Gamma(n\alpha + \alpha(\lambda+1))}.
\end{equation}

Using the following relations
\begin{equation}
	C_n^{(\lambda)}(1) = \frac{\Gamma(n+2\lambda)}{\Gamma(n+1)\Gamma(2\lambda)}
	\quad \text{and} \quad
	N_n^2 = \frac{n!(n+\lambda)\Gamma^2(\lambda)}{\pi\, 2^{1-2\lambda}\Gamma(n+2\lambda)},
\end{equation}

we obtain
\begin{equation}
	\left[N_n C_n^{(\lambda)}(1)\right]^2 = \frac{(n+\lambda)\Gamma^2(\lambda)\Gamma(n+2\lambda)}{\pi\, 2^{1-2\lambda}\, n!\, \Gamma^2(2\lambda)}.
\end{equation}

Applying the duplication formula of Gamma function \cite{55}
\begin{equation}
	\Gamma(2\lambda) = \frac{2^{2\lambda-1}}{\sqrt{\pi}} \Gamma(\lambda)\, \Gamma\left(\lambda + \frac{1}{2}\right),
\end{equation}

Equation (90) is reduced into
\begin{equation}
	\left[N_n C_n^{(\lambda)}(1)\right]^2 = \frac{(n+\lambda)\Gamma(n+2\lambda)}{n!\,\Gamma(2\lambda)\,\Gamma\left(\lambda+\frac{1}{2}\right)} \frac{1}{\sqrt{\pi}}.
\end{equation}

Using the asymptotic expansion of a ratio of Gamma functions for large $\lambda$, such that
\begin{equation}
	\frac{\Gamma(z+a)}{\Gamma(z+b)} \sim z^{a-b}, \quad z \to \infty,
\end{equation}

we have
\begin{equation}
	\frac{\Gamma(n+2\lambda)}{\Gamma(2\lambda)} \sim (2\lambda)^n
	\quad \text{and} \quad
	\Gamma\left(\lambda + \frac{1}{2}\right) \sim \lambda^{-1/2}\Gamma(\lambda), \quad \lambda \to \infty.
\end{equation}

Consequently,
\begin{equation}
	\left[N_n C_n^{(\lambda)}(1)\right]^2 \sim \frac{2^n}{n!\sqrt{\pi}}\, \lambda^{n+\frac{1}{2}},
\end{equation}
Consequently,
\begin{equation}
	\left[N_n C_n^{(\lambda)}(1)\right]^2 \sim \frac{2^n}{n!\sqrt{\pi}}\, \lambda^{n+\frac{1}{2}},
\end{equation}
\begin{equation}
	\left[N_n C_n^{(\lambda)}(1)\right]^{2\alpha} \sim \left(\frac{2^n}{n!\sqrt{\pi}}\right)^\alpha \lambda^{\alpha\left(n+\frac{1}{2}\right)}.
\end{equation}

Using the relation (93), we obtain
\begin{equation}
	\frac{\Gamma\left(\alpha(\lambda+1) - \frac{1}{2}\right)}{\Gamma(n\alpha + \alpha(\lambda+1))} \sim (\alpha\lambda)^{-n\alpha - \frac{1}{2}}.
\end{equation}

Substituting these asymptotic estimates into Eq.~(88), we find
\begin{equation}
	\mathcal{W}^{(\alpha)}[\rho_n^\lambda(x)] \sim \frac{1}{\sqrt{\tau}} \left(\frac{2^n}{n!\sqrt{\pi}}\right)^\alpha \Gamma\left(n\alpha + \frac{1}{2}\right) \lambda^{\alpha\left(n+\frac{1}{2}\right)} (\alpha\lambda)^{-n\alpha - \frac{1}{2}}.
\end{equation}

After simplification,
\begin{equation}
	\mathcal{W}^{(\alpha)}[\rho_n^\lambda(x)] \sim \frac{1}{\sqrt{\tau}} \frac{2^{n\alpha}\, \Gamma\left(n\alpha + \frac{1}{2}\right)}{(n!)^\alpha\, \pi^{\alpha/2}\, \alpha^{n\alpha + \frac{1}{2}}}\, \lambda^{\frac{\alpha-1}{2}}.
\end{equation}
\section*{Appendix C: Numerical values of Heisenberg-like uncertainty relation \eqref{Hei} }\label{hu}

\section*{Appendix D: Numerical values of R\'enyi and Tsallis information uncertainty relations}\label{vb}
\begin{table}[h]
	\centering
	\caption{Heisenberg-like uncertainty $\tau = 0.1$}
	\begin{tabular}{cccc}
		\hline\hline
		$n$ & $\sigma_x^2$ & $\sigma_p^2$ & $\sigma_x^2 \sigma_p^2$ \\
		\hline
		0 & 0.000  & 0.710    & 0.000     \\
		1 & 5.000  & 17.637   & 88.186    \\
		2 & 8.333  & 124.651  & 1038.755  \\
		3 & 11.250 & 499.718  & 5621.824  \\
		4 & 14.000 & 1486.958 & 20817.406 \\
		5 & 16.667 & 3734.548 & 62242.464 \\
		\hline
	\end{tabular}
\end{table}

\begin{table}[h]
	\centering
	\caption{Heisenberg-like uncertainty for $\tau = 0.5$}
	\begin{tabular}{cccc}
		\hline\hline
		$n$ & $\sigma_x^2$ & $\sigma_p^2$ & $\sigma_x^2 \sigma_p^2$ \\
		\hline
		0 & 0.000 & 1.587    & 0.000     \\
		1 & 1.000 & 39.438   & 39.438    \\
		2 & 1.667 & 278.727  & 464.546   \\
		3 & 2.250 & 1117.403 & 2514.156  \\
		4 & 2.800 & 3324.938 & 9309.827  \\
		5 & 3.333 & 8350.703 & 27835.676 \\
		\hline
	\end{tabular}
\end{table}

\begin{table}
	\centering
	\caption{Numerical values of the Rényi information uncertainty relation $\tau=0.1$.}
	\label{tab2}
	\begin{tabular}{rrrrrrrr}
			\hline\hline
		$\tau$ &    $\lambda$ &    $\alpha$ &     $\beta$ &  $n$ & $ \mathcal{R}^{(\alpha)}[\rho_n^\lambda(x)]$ &  $\mathcal{R}^{(\beta)}[\rho_n^\lambda(p)] $ &  $   \Delta\left[\mathcal{R}^{(\alpha)}  \right]$ \\
		\hline
		0.100000 & 10.512492 & 0.600000 & 3.000000 &  0 &    1.232796 &   0.827018 & 0.001899 \\
		0.100000 & 10.512492 & 0.600000 & 3.000000 &  1 &    1.526347 &   1.176765 & 0.645197 \\
		0.100000 & 10.512492 & 0.600000 & 3.000000 &  2 &    1.713435 &   1.383719 & 1.039240 \\
		0.100000 & 10.512492 & 0.600000 & 3.000000 &  3 &    1.854905 &   1.535973 & 1.332963 \\
		0.100000 & 10.512492 & 0.600000 & 3.000000 &  4 &    1.969854 &   1.655061 & 1.567000 \\
		0.100000 & 10.512492 & 0.600000 & 3.000000 &  5 &    2.067040 &   1.753746 & 1.762872 \\
		0.100000 & 10.512492 & 1.500000 & 0.750000 &  0 &    0.963762 &   1.163921 & 0.002124 \\
		0.100000 & 10.512492 & 1.500000 & 0.750000 &  1 &    1.288057 &   1.470680 & 0.633178 \\
		0.100000 & 10.512492 & 1.500000 & 0.750000 &  2 &    1.488060 &   1.661038 & 1.023539 \\
		0.100000 & 10.512492 & 1.500000 & 0.750000 &  3 &    1.636617 &   1.801291 & 1.312349 \\
		0.100000 & 10.512492 & 1.500000 & 0.750000 &  4 &    1.755189 &   1.915938 & 1.545568 \\
		0.100000 & 10.512492 & 1.500000 & 0.750000 &  5 &    1.853457 &   2.012024 & 1.739923 \\
		0.100000 & 10.512492 & 2.000000 & 0.666667 &  0 &    0.899085 &   1.202128 & 0.001711 \\
		0.100000 & 10.512492 & 2.000000 & 0.666667 &  1 &    1.229281 &   1.502719 & 0.632499 \\
		0.100000 & 10.512492 & 2.000000 & 0.666667 &  2 &    1.431715 &   1.690899 & 1.023113 \\
		0.100000 & 10.512492 & 2.000000 & 0.666667 &  3 &    1.581525 &   1.829047 & 1.311070 \\
		0.100000 & 10.512492 & 2.000000 & 0.666667 &  4 &    1.700371 &   1.942832 & 1.543702 \\
		0.100000 & 10.512492 & 2.000000 & 0.666667 &  5 &    1.798014 &   2.038240 & 1.736754 \\
		0.100000 & 10.512492 & 2.500000 & 0.625000 &  0 &    0.854497 &   1.223757 & 0.001422 \\
		0.100000 & 10.512492 & 2.500000 & 0.625000 &  1 &    1.188338 &   1.520766 & 0.632272 \\
		0.100000 & 10.512492 & 2.500000 & 0.625000 &  2 &    1.392267 &   1.707695 & 1.023130 \\
		0.100000 & 10.512492 & 2.500000 & 0.625000 &  3 &    1.542806 &   1.844600 & 1.310574 \\
		0.100000 & 10.512492 & 2.500000 & 0.625000 &  4 &    1.661587 &   1.957866 & 1.542621 \\
		0.100000 & 10.512492 & 2.500000 & 0.625000 &  5 &    1.758364 &   2.052881 & 1.734413 \\
		0.100000 & 10.512492 & 3.000000 & 0.600000 &  0 &    0.821434 &   1.237693 & 0.001213 \\
		0.100000 & 10.512492 & 3.000000 & 0.600000 &  1 &    1.157746 &   1.532363 & 0.632194 \\
		0.100000 & 10.512492 & 3.000000 & 0.600000 &  2 &    1.362687 &   1.718477 & 1.023249 \\
		0.100000 & 10.512492 & 3.000000 & 0.600000 &  3 &    1.513683 &   1.854566 & 1.310335 \\
		0.100000 & 10.512492 & 3.000000 & 0.600000 &  4 &    1.632222 &   1.967485 & 1.541791 \\
		0.100000 & 10.512492 & 3.000000 & 0.600000 &  5 &    1.728007 &   2.062242 & 1.732334 \\
		\hline
	\end{tabular}
\end{table}

\begin{table}
	\centering
	\caption{Nmerical values of the Tsallis information uncertainty relation for fixed $\tau=0.1$.}
	\label{tab}
	\begin{tabular}{rrrrrrrr}
		\hline\hline
		$\tau$ &    $\lambda$ &    $\alpha$ &    $ \beta$ &  $n$ &   $\mathcal{T}^{(\alpha)}[\rho_n^\lambda(x)]$ &  $\mathcal{T}^{(\beta)}[\rho_n^\lambda(p)]$ &   $\Delta\left[\mathcal{T}^{(\alpha)}  \right] $      \\
		\hline
0.100000 & 10.512492 & 0.600000 & 3.000000 &  0 & 1.593536 &0.404362& 0.000000 \\
0.100000 & 10.512492 & 0.600000 & 3.000000 &  1 &2.103558 &  0.452483& 0.207411 \\
0.100000 & 10.512492 & 0.600000 & 3.000000 &  2 &2.461286 &0.468589 &  0.314311 \\
0.100000 & 10.512492 & 0.600000 & 3.000000 &  3 &2.750129 & 0.476835 & 0.390756 \\
0.100000 & 10.512492 & 0.600000 & 3.000000 &  4 &2.997163 &0.481744 &  0.451918 \\
0.100000 & 10.512492 & 0.600000 & 3.000000 &  5 &3.215072 &0.485014 &  0.505241 \\

0.100000 & 10.512492 & 1.500000 & 0.750000 &  0 & 0.764759&1.350953 &  0.000005 \\
0.100000 & 10.512492 & 1.500000 & 0.750000 &  1 &0.949655 &1.777461 & 0.155396 \\
0.100000 & 10.512492 & 1.500000 & 0.750000 &  2 &1.049610 &2.059055 & 0.243464 \\
0.100000 & 10.512492 & 1.500000 & 0.750000 &  3 &1.117646 &2.275274 &  0.306986 \\
0.100000 & 10.512492 & 1.500000 & 0.750000 &  4 & 1.168436&2.457737 & 0.357581 \\
0.100000 & 10.512492 & 1.500000 & 0.750000 &  5 &1.208307 &2.614740 & 0.400100 \\
0.100000 & 10.512492 & 2.000000 & 0.666667 &  0 &0.593058 &1.478649 & 0.000005 \\
0.100000 & 10.512492 & 2.000000 & 0.666667 &  1 & 0.707497 & 1.950649 & 0.103504 \\
0.100000 & 10.512492 & 2.000000 & 0.666667 &  2 & 0.761101 &2.271134 &0.163486 \\
0.100000 & 10.512492 & 2.000000 & 0.666667 &  3 &0.794339 &2.519540 &  0.206847 \\
0.100000 & 10.512492 & 2.000000 & 0.666667 &  4 &0.817384 & 2.732908 &0.241277 \\
0.100000 & 10.512492 & 2.000000 & 0.666667 &  5 &0.834373 &2.918161 & 0.270076 \\
0.100000 & 10.512492 & 2.500000 & 0.625000 &  0 & 0.481632 &1.552923 &  0.000003 \\
0.100000 & 10.512492 & 2.500000 & 0.625000 &  1 &0.554523 &2.050067 & 0.051718 \\
0.100000 & 10.512492 & 2.500000 & 0.625000 &  2 & 0.584077 &2.392566 & 0.081932 \\
0.100000 & 10.512492 & 2.500000 & 0.625000 &  3 & 0.600770 &2.659087 & 0.103778 \\
0.100000 & 10.512492 & 2.500000 & 0.625000 &  4 &0.611525 &2.890169 &  0.121088 \\
0.100000 & 10.512492 & 2.500000 & 0.625000 &  5 &0.618976 &3.091732 & 0.135535 \\
0.100000 & 10.512492 & 3.000000 & 0.600000 &  0 & 0.403288 &1.601563& 0.000036 \\
0.100000 & 10.512492 & 3.000000 & 0.600000 &  1 & 0.450641 &2.114649 &0.550606 \\
0.100000 & 10.512492 & 3.000000 & 0.600000 &  2 & 0.467239 &2.471301 & 0.873057 \\
0.100000 & 10.512492 & 3.000000 & 0.600000 &  3 & 0.475778 &2.749419 & 1.106266 \\
0.100000 & 10.512492 & 3.000000 & 0.600000 &  4 & 0.480891 & 2.991957 & 1.290914 \\
0.100000 & 10.512492 & 3.000000 & 0.600000 &  5 & 0.484222 &3.204113 &1.444992 \\
		\hline
	\end{tabular}
\end{table}

\end{document}